\documentclass[structabstract]{aa}
\pdfoutput=1

\usepackage{graphicx}
\usepackage{natbib}
\usepackage{textcomp}
\usepackage{amsmath}
\usepackage{amssymb}
\usepackage{wasysym}
\usepackage{fancybox}
\usepackage{multirow}

\begin{document}

\title{Long-term stability of spotted regions and the activity-induced Rossiter-McLaughlin effect on V889 Herculis}
\subtitle{A synergy of photometry, radial velocity measurements, and Doppler imaging}

\author{K. F. Huber\inst{1}
  \and U. Wolter\inst{1}
  \and S. Czesla\inst{1}
  \and J. H. M. M. Schmitt\inst{1}
  \and M. Esposito\inst{1,2}
  \and I. Ilyin\inst{3}
  \and \mbox{J. N. Gonz\'alez-P\'erez\inst{1}}}

\institute{Hamburger Sternwarte, Universit\"at Hamburg, Gojenbergsweg 112, 21029 Hamburg, Germany
           \and Th\"uringer Landessternwarte Tautenburg, Sternwarte 5, 07778 Tautenburg, Germany
           \and Astrophysikalisches Institut Postdam, An der Sternwarte 16, 14482 Potsdam, Germany}

\date{Received ... / Accepted ... }

% \abstract {} {Text of aims} {Text of methods} {Text of results} {}
\abstract {The young active G-dwarf star V889 Herculis (HD 171488) shows pronounced spots in Doppler images as well as large variations in photometry and radial velocity (RV) measurements. However, the lifetime and evolution of its active regions are not well known.}
          {We study the existence and stability of active regions on the star's surface using complementary data and methods. Furthermore, we analyze the correlation of spot-induced RV variations and Doppler images.}
          {Photometry and high-resolution spectroscopy are used to examine stellar activity. A CLEAN-like Doppler Imaging (DI) algorithm is used to derive surface reconstructions. We study high-precision RV curves to determine their modulation due to stellar activity in analogy to the Rossiter-McLaughlin effect. To this end we develop a measure for the shift of a line's center and compare it to RV measurements.}
          {We show that large spotted regions exist on V889 Her for more than one year remaining similar in their large scale structure and position. This applies to several time periods of our observations, which cover more than a decade.
          Furthermore we use DI line profile reconstructions to identify influences of long-lasting starspots on RV measurements.
          In this way we verify the RV curve's agreement with our Doppler images.
          Based on long-term RV data we confirm V889 Her's rotation period of $1.3371 \, \pm \, 0.0002$~days.}
          {}


\titlerunning{Long-term stability of spotted regions and the activity-induced Rossiter-McL. effect}
\maketitle

%%%%%%%%%%%%%%%%%%%%%%%%%%%%%%%%%%%%%%%%%%%%%%%%%%%%%%%%
\section{Introduction}
\label{Sec:Intro}

The Sun is the only star which is sufficiently close to resolve surface inhomogeneities in detail.
Most prominent among them are sunspots.
At solar maximum they cover approximately $0.3 \, - \, 0.4$~\% of the total solar surface \citep{Solanki2004} and individual sunspots reach diameters of roughly $2$ degrees \citep{Schrijver2002} on average.
Two examples of exceptionally large spots on the Sun were observed in March 1947 \citep{Willis1979}, covering about $0.6$~\% of the solar disk, and on March 29, 2001 \citep{Koso2001}, with a size of more than $0.35$~\% of the visible surface.
The appearance and movement of sunspots is often displayed in `butterfly diagrams' which show that their positions are confined to a band of $\pm \, 30 \, - \, 40$ degrees above and below the equator (e.g. \citealt{Li2001}).
Their lifetimes vary from a few days to several weeks \mbox{($\apprle$~month)} and appear
to be related to the spot-size, larger spots persist longer \citep{Petro1997}.

This picture seems to be different in several respects from other stars of similar spectral type, although these observations are biased towards stars with small rotation periods, which is a basic requirement of the currently available techniques to study stellar surface inhomogeneities.
Contrary to the slowly rotating Sun we observe that many rapid rotators \mbox{($v\sin(i) \, \apprge \, 25$~km/s)} possess large areas covered with features of significantly lower temperatures.
One of the most extreme examples was found in the binary system VW Cep with a surface coverage of $\sim \, 70$~\% \citep{Hendry2000}.
Starspots or spot groups with sizes of several dozens of degrees have been detected not only at low and intermediate latitudes, but also covering the polar regions.
These polar spots form long-lasting features, sometimes persisting over years or possibly even decades \citep{Vogt1999,Jeffers2007}.
The limits of current techniques often prevent us from discerning large monolithic spots from groups of individual spots within an active region.
Whatever their detailed nature, these structures have much longer lifetimes than sunspots.
Large spotted regions have been monitored on some stars for years showing only small changes (e.g. \citealt{Korhonen2007} and \citealt{Lanza2006}),
although individual starspots appear to evolve on timescales of several weeks
\citep{Barnes1998, Wolter2005a}.

Such observations gave rise to the idea of `active longitudes'; regions on the stellar surface that persist in their large scale structure \citep{Korhonen2001}.
The term `active longitudes' may partially reflect the weakness of surface reconstruction methods to resolve the positions of stellar features in longitude more accurately than in latitude.
Such regions show a high spot coverage stretching over a large area of the star, persisting for many months or years, which translates into several hundreds of stellar rotations.
It is largely unknown whether their internal structure is variable, i.e., to what degree spots move, disappear, or form inside of them.
However, in some cases their global structure remains stable over large time periods (e.g. \citealt{Berdy2005}).

In this paper we show that large spotted regions on V889's surface persist on time scales of hundreds of days.
To this end we use photometric data, Doppler images, and radial velocity (RV) measurements.
Section~\ref{Sec:Object} summarizes characteristics of the star and its surface features.
In Sect.~\ref{Sec:Obs} we present observational characteristics of the Hipparcos data, the Doppler Imaging (DI) data, and the RV measurements.
Section~\ref{Sec:LCanalysis} explains the process of the lightcurve analysis, while Sects.~\ref{Sec:RV} and~\ref{Sec:DI} contain the analysis of the RV measurements and the DI process.
Section~\ref{Sec:IntDis} gives a discussion on all data sets and their correlation,
followed by a summary of the main results in Sect.~\ref{Sec:Summary}.
% Finally, the appendix covers the discussion \textbf{on how parameters of our light curve models may be correlated}, the
% mathematical derivation and discussion of the equation used to calculate RV shifts from the DI results,
% and a representative sample of DI line profile reconstructions.
Finally, the appendix covers mathematical aspects
%the discussion on the mathematical derivation and discussion
of the equation used to calculate RV shifts from line profiles,
%the DI results,
and contains all DI line profile reconstructions.

%%%%%%%%%%%%%%%%%%%%%%%%%%%%%%%%%%%%%%%%%%%%%%%%%%%%%%%%
\section{Object}
\label{Sec:Object}

\begin{table}[t!]
  \begin{minipage}[h]{0.5\textwidth}
    \renewcommand{\footnoterule}{}
    \caption{Stellar parameters of V889 Her \label{Tab:V889prop}}
    \begin{center}
      \begin{tabular}{l c}
      \hline \hline
      Parameter \hspace*{3.5cm} & Value \\
      \hline
      Spectral type\footnote{SIMBAD Astronomical database}
                           & G0V \\
      Distance (Hipparcos)\footnote{Hipparcos Catalogue, \citet{Perryman1997}}
                           & $37.2 \pm 1.2$ pc \\
      $M_{\mathrm{hipp, \, mean}}$\footnote{mean Hipparcos magnitude and standard error}
                           & $7\fm523 \, \pm \, 0\fm004$ \\
      $M_{\mathrm{hipp, \, max}}$\footnote{maximum Hipparcos magnitude}
                           & $7\fm48 \, \pm \, 0\fm01$ \\
      $M_{\mathrm{hipp, \, min}}$\footnote{minimum Hipparcos magnitude}
                           & $7\fm57 \, \pm \, 0\fm01$ \\
      $V_{\mathrm{max}}$\footnote{\citet{Strassmeier2003}}
                           & $7\fm34$ \\
      $v \sin(i)^f$
                           & $39.0 \pm 0.5 \ \mathrm{km} \, \mathrm{s}^{-1}$ \\
      Rotation period$^f$
                           & $1.3371 \pm 0.0002$ d \\
      Inclination $i$ $^f$ & $\sim \, 55$\textdegree \\
      \hline
      \end{tabular}
    \end{center}
  \end{minipage}
\end{table}

The fast-rotating solar analog V889 Herculis \mbox{(HD 171488)} has been a subject of astrophysical studies for many years.
Its good visibility from the northern hemisphere, large brightness, pronounced photometric variability, and high rotation velocity make it an interesting target.
Especially its rotation period of $1.3$~days is convenient to study the lifetimes of atmospheric structures, since observations covering four consecutive nights offer a complete coverage of all rotation phases.
Therefore, it is a well-suited candidate for DI and other studies of stellar activity.
Doppler images of V889 Her were published by \citet{Strassmeier2003} and \citet{Marsden2006}.
Some of its basic properties are summarized in Table~\ref{Tab:V889prop}.

Surface reconstructions of V889 Her from these authors reveal a large polar spot that changes in size and structure but has yet always been present. Additionally, all published Doppler images show some medium- and low-latitude spots which may stretch over a few dozens of degrees and, thus, cover a significant fraction of the stellar surface.
\citet{Strassmeier2003} present surface maps dominated by an asymmetric polar spot with a temperature difference of \mbox{$\Delta T \, \approx \, 1 \, 600$~K} to the unspotted surface at \mbox{$\approx \, 5 \, 900$~K}.
Several low-latitude spots are resolved as well with \mbox{$\Delta T \, \approx \, 500-800$~K}.
\citeauthor{Strassmeier2003} determine a rotation period of \mbox{$P \, = \, 1.3371 \, \pm \, 0.0002$~days} using long-term photometry.
\citet{Marsden2006} reconstruct a polar spot of similar size down to almost +60\textdegree \ in latitude and two pronounced low-latitude spots at +30\textdegree.
Additionally reconstructions for the magnetic field topology were possible by Zeeman Doppler Imaging (ZDI), which yield only weak signs of a polar spot (in agreement with ZDI images of other objects, see e.g. \citealt{Donati2003}) and confirm the low-latitude features. \citet{Marsden2006} find a different rotation period of  \mbox{$P \, = \, 1.313 \, \pm \, 0.004$~days} and a differential rotation of \mbox{$\alpha \, = \, \Delta P/P \, = \, 0.08$}, while for the Sun a value of \mbox{$\alpha \, \approx \, 0.2\, - \, 0.3$} is measured.
$\Delta P$ denotes the difference between the equatorial and the polar rotation period.

Up to now it has not been known on which time scales spotted regions on the surface of V889 Her form and dissolve and how heavily the effects of differential rotation and meridional flows influence their development in shape and position.
All surface features are presumably in constant evolution, but there are indications in previous Doppler images and reconstructions in this paper that especially large active regions besides the polar spot may have fixed positions and high stability.
Evidence of such preferred longitudes was detected before in other stars (e.g. \citealt{Berdy2007}), 
while
%Note, however, that
\citet{Barnes1998} do not find long-lived spots on rapidly rotating G dwarfs indicating that this is not a general characteristic for this type of stars.

%%%%%%%%%%%%%%%%%%%%%%%%%%%%%%%%%%%%%%%%%%%%%%%%%%%%%%%%
\section{Observations}
\label{Sec:Obs}

\begin{figure}[b!]
   \centering
   \includegraphics[width=0.47\textwidth]{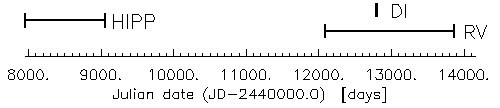}
\caption{Time coverage of observations used in this paper. HIPP represents the Hipparcos, DI the Doppler Imaging, and RV the radial velocity data sets.
\label{Fig:TLall}}
\end{figure}

We use three different kinds of data sets to determine the stability of spotted regions on V889 Her: long-term photometric observations from the Hipparcos mission, and two different time series of high-resolution spectra used for radial velocity (RV) measurements and Doppler Imaging.
The short-term time series suitable for DI yield snapshots of the spot distribution, which we correlate with our long-term RV measurements covering the same time period.
Figure~\ref{Fig:TLall} visualizes the time coverage of the three data sets used in this work.

\subsection{Hipparcos data}
\label{Sec:OBShipp}

The Hipparcos satellite obtained a lightcurve of V889 Her during its 3.3~years mission.
The available data cover the time between March 1990 and March 1993 with about 150 unevenly sampled photometric data points \citep{Perryman1997}.
% More detailed information on this data set is given in Table~\ref{Tab:HIPPdata}.

\subsection{RV measurements}
\label{Sec:OBSrv}

The RV observations were carried out at the \mbox{2 m} Alfred Jensch Telescope at the
Th\"uringer Landessternwarte in Tautenburg, Germany.
It is equipped with a Coud\'e Echelle spectrograph, which provides a resolving power of \mbox{$R = 67 \, 000$}
and a wavelength range of \mbox{$4 \, 700$ - $7 \, 400$~\AA}.
In total, $62$~radial velocity shifts were analyzed in this paper.
For the purpose of a stable wavelength calibration, an iodine cell was used \citep{Marcy1992};
for more information about the spectrograph see \citet{Hatzes2005}.

The exposure times vary between 15 and 20 minutes, the typical signal-to-noise ratio is 70.
The data was reduced with the Image Reduction and Analysis Facility (IRAF, \citealt{Tody1993}).
For the RV measurements the spectral range between $5 \, 000$~\AA \ and $6 \, 300$~\AA \ is split up into about 130 pieces (`chunks').
RV shifts are determined independently for each wavelength piece, the mean of all pieces yields the radial velocity with its error given as the standard deviation.

For the analysis, the RV curves are split up into shorter subintervals.
% the same way as described for the photometric light curves in Sect.~\ref{Sec:LCanalysis}.
Their exact definitions and an overview of the entire observations are given in Fig.~\ref{Fig:RVdata}.

\subsection{Doppler Imaging spectra}
\label{Sec:OBSdi}

The time series of high-resolution spectra used for Doppler Imaging was observed in June 2003
at the Nordic Optical Telescope (NOT) on La Palma.
A total of 61 spectra of V889 Her were obtained with the SOFIN Echelle spectrograph mounted at the Cassegrain focus of the \mbox{2.56 m} telescope, covering several preselected spectral regions between $3 \, 750$~\AA \ and $11 \, 300$~\AA \
with a width of e.g. 60~\AA \ per region around 6000~\AA.
The average signal-to-noise ratio of these spectra is above 150.
The chosen slit width of 65~$\mu$ provides a nominal resolving power of $76 \, 000$.
The data cover 8 nights (2 blocks of 4 consecutive nights), i.e., two full rotations of the star, with a homogeneous phase coverage.
Approximately 3 rotations of the star lie between the two obtained Doppler images.
The average exposure time for these observations was 20 minutes.
A short summary of the observation dates and the number of spectra is given in Table~\ref{Tab:NOTdata}.

% The Echelle spectra were reduced with the software package described in \citet{IlyinPhD}.
% This includes standard steps of bias subtraction, master flat field correction, scattered light subtraction approximated with bi-cubic splines, weighted extraction of spectra with cosmic spikes rejection, and accurate wavelength calibration with the use of two Th-Ar spectra taken before and after each object exposure.
% Continuum rectification and CCD fringes removal was done with the flat field spectra reduced in a similar way.

The Echelle spectra were reduced with the software package described in \citet{IlyinPhD},
performing
%This includes standard steps of 
bias subtraction, master flat field correction, scattered light modeling with
bi-cubic splines, and optimum extraction of spectra including cosmic spikes rejection.
Blaze correction and CCD fringe removal was accomplished using flat field exposures that were reduced in a similar way.
The wavelength calibration is based on two Th-Ar spectra taken before and after each object exposure.

\begin{table}[t!]
  \begin{minipage}[h]{0.5\textwidth}
  \renewcommand{\footnoterule}{}
  \caption{Nordic Optical Telescope data
    \label{Tab:NOTdata}}
\begin{center}
    \begin{tabular}{l c c c c}
      \hline \hline
      N\footnote{number of nights} & Observation dates & Julian date\footnote{JD - $\mathrm{JD}_0$ with $\mathrm{JD}_0 = 2 \, 449 \, 000.0$ days}
                                   & Total\footnote{total number of spectra available}
                                   & Used\footnote{number of spectra used for DI} \\
      \hline
      \multirow{2}{.8cm}{1 - 4} & June 09, 2003 & 3799.54 & \multirow{2}{.5cm}{42}
    & \multirow{2}{.4cm}{39}\footnote{three spectra were dismissed due to poor SNR,
                                      \mbox{\hspace*{.8cm} see Sect.~\ref{Sec:DIclean}}} \\
                                & June 12, 2003 & 3802.75 & & \\
      \hline
      \multirow{2}{.8cm}{9 - 12} & June 16, 2003 & 3807.47
                                & \multirow{2}{.5cm}{19} & \multirow{2}{.4cm}{19} \\
                                & June 20, 2003 & 3810.69 & & \\
      \hline
    \end{tabular}
\end{center}
  \end{minipage}
\end{table}

%%%%%%%%%%%%%%%%%%%%%%%%%%%%%%%%%%%%%%%%%%%%%%%%%%%%%%%%%%%%%
\section{Lightcurve analysis}
\label{Sec:LCanalysis}

  \subsection{Our modeling approach}
   Lightcurve modeling has a long-standing history in astronomy. An overview on its history as well as on
   different approaches towards a solution of this problem can for instance be found in \citet{Eker1994}.
   
   The Hipparcos data are of limited accuracy and
   %the measurements are, furthermore,
   rather inhomogeneously sampled in time.
   Therefore, we revert to a simple but robust modeling approach. We subdivide the star into homogeneously spaced
   %`longitudinal strips', i.e., into areas included within a specific
   longitude intervals, and assume that these areas are homogeneously covered by spots.
   In the fitting process the relative brightness (spot coverage) of these
   strips is then varied until the observed lightcurve is matched optimally.
   We apply two different fit procedures:
   Either we allow fits
   restricted by a specific assumption about the number of active regions on the stellar surface; in this case every
   active region is modeled by a %brightness decrease with a
   longitudinal Gaussian brightness distribution.
   Or, alternatively, a `free' fit
   can be carried out, where all surface elements are varied independently.
   
%    \textbf{Note that when the number of fitted data points equals the number of longitude intervals (which may also be achieved
%    by combining several data points), this fitting approach
%    resembles a matrix inversion, which is unique (and defined given a maximal rank of the matrix, which will usually be
%    satisfied). Therefore, the outcomes produced by the fit are also unique, within the limits
%    set by the measurement errors.}
   
   Note that each individual longitude interval produces the same lightcurve profile (normalized by its intensity, i.e.
   spot coverage), phase shifted according to its position on the surface. The resulting lightcurve is a linear superposition of the
   individual contributions. When we use a relatively low number of longitude intervals (say $10$), it is quite obvious that the 
   contributions are linearly independent (given a non pole-on view), and the superposition is uniquely defined. Working with
   real data, measurement errors and phase coverage limit the uniqueness of the reconstruction; nevertheless, with a low number
   of longitude intervals, only neighboring intervals show considerable dependence.
   
%    Regularization of this model does not seem necessary since we intent to reproduce longitudes of time-averaged
%    activity regions with a relatively rough re\-solution and not shapes and sizes of physical spots.

   In either case latitudinal information is not recovered, which $-$ given the data quality and sampling $-$
   would be an ambitious task anyway.

  \subsection{Lifetime analysis of structures imprinted on the Hipparcos lightcurve}
  
  \begin{figure}[t]
    \includegraphics[angle=-90,width=0.5\textwidth]{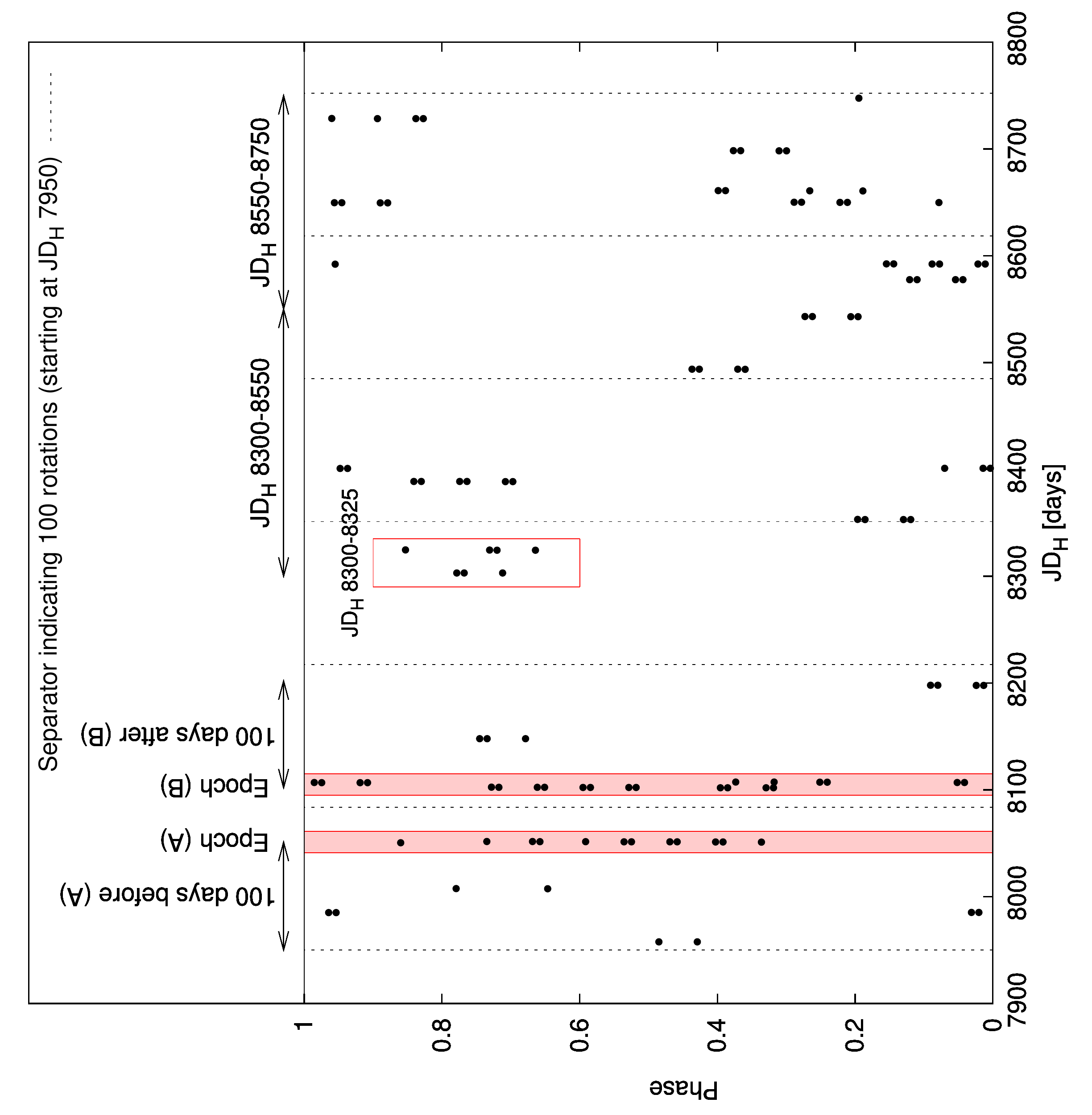}
    \caption{Time vs. phase diagram \mbox{(period$\, = \, 1.3371$~days)} for the Hipparcos measurements (represented by points).
    Periods covering $100$ rotations are indicated by vertical, dotted lines.
    Time and phase intervals of special interest are marked by arrows and boxes; see text for more details.
    \label{fig:HippDataInfo}}
  \end{figure}
  
\begin{figure}[t]
    \includegraphics[angle=-90,width=0.5\textwidth,clip=,bb=30 0 740 720]{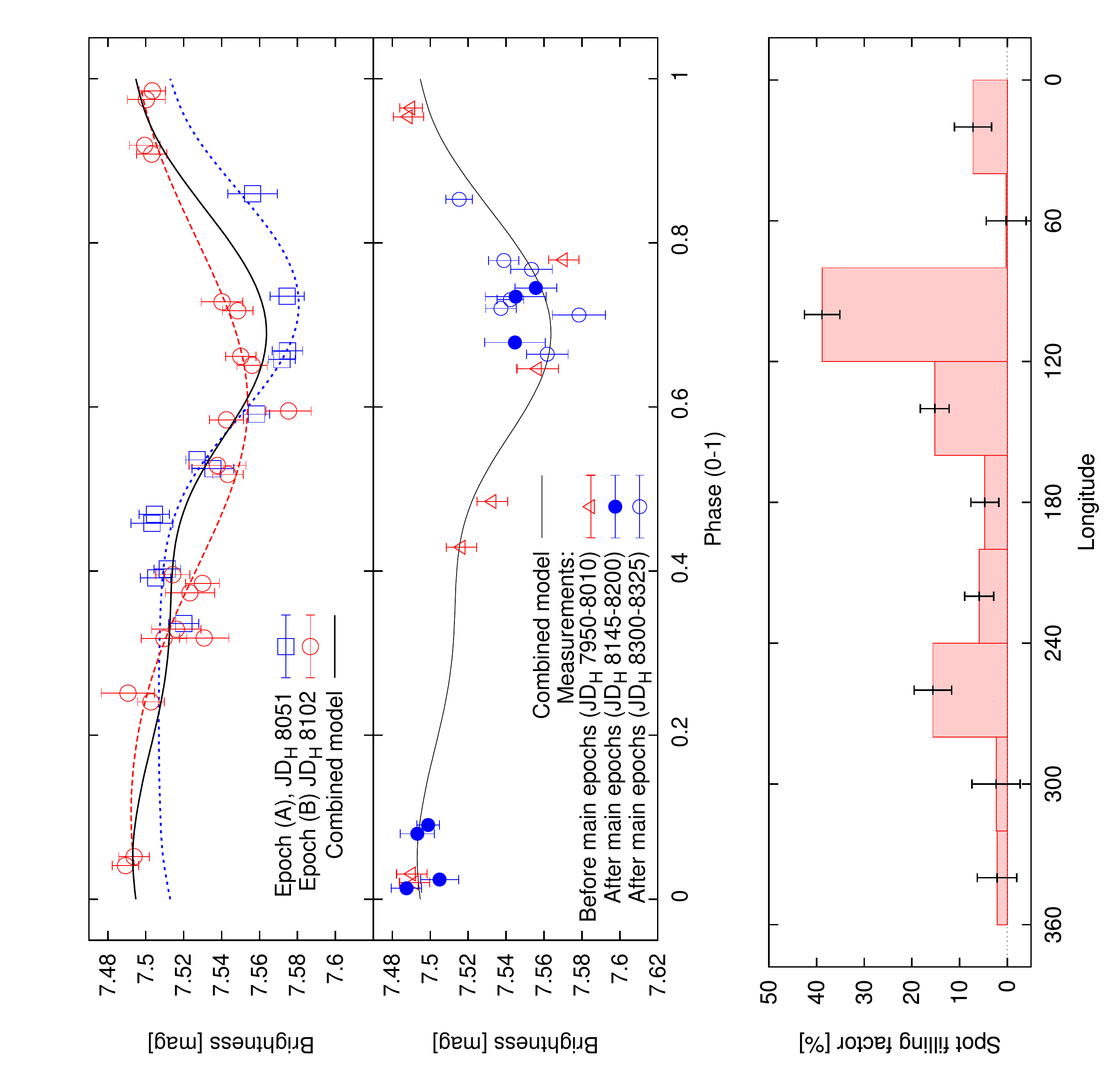}
    \caption{\textit{Upper panel:} The data and models for observation epoch~(A) (squares) and (B) (circles).
    The dotted and dashed lines indicate model fits obtained for epoch~(A) and (B) assuming a surface dominated
    by one active region, respectively. The solid line constitutes a model based on the combined data of epoch~(A) and (B).
    \textit{Middle panel:} The `combined' model from the upper panel as a solid line as well as additional data points
    observed before and after epochs~(A) and (B).
    \textit{Lower panel:} The spot filling factor obtained from longitude reconstruction.
    The spot-photosphere contrast was assumed to be $50$~\%.
    (Note that the error bars do not take into account mutual dependencies of the bins.)
    \label{fig:HippABD}}
  \end{figure}

  As demonstrated in Fig.~\ref{fig:HippDataInfo}, the sampling of the Hipparcos lightcurve is inhomogeneous
  so that the data contain quite a number of isolated measurements.
  Fortunately, there are some observation epochs with much better phase coverage. The
  first one (A) can be found at $\mathrm{JD}_H$\footnote{$\mathrm{JD}_H\, =\, \mathrm{JD}\, -\, 2\,440\,000$~days}
  \mbox{$\approx\, 8050$} where we find $12$~data points observed during about a day,
  and the second one (B) is found about $50$~days later containing $22$~measurements distributed over $5$~days
  (cf. Fig.~\ref{fig:HippDataInfo}).
  The corresponding data points are shown in the upper panel of Fig.~\ref{fig:HippABD}, where we also
  present lightcurve reconstructions obtained under the assumption of a single active region on the stellar surface.
  The measurements originating
  from epoch~(A) cover only phases \mbox{$0.3\, -\, 0.9$}, whereas epoch~(B) provides nearly full
  phase coverage.
  Clearly, the minimum in interval (A) is sharper than in (B) and also displaced in phase by about $0.1$ corresponding to
  a longitude shift of $\approx 36$\textdegree~for the activity center. While this shows that there is significant evolution of the lightcurve
  within $50$~days, the lightcurve also maintains its overall appearance showing a pronounced minimum at about $120$\textdegree~longitude
  as the most distinct feature.
  The differences between the two epochs can be caused by evolution of the stellar surface and/or
  by an inappropriate lightcurve folding due to an incorrect rotation period.
  According to \citet{Marsden2006}, V889~Her shows strong differential
  rotation making its period a function of latitude; folding the data with a period inconsistent with the location
  of the active regions could mimic surface evolution. Note that adopting a rotation
  period of \mbox{$1.3371\, \pm\, 0.0002$~days}, as given by \citet{Strassmeier2003}, the relative phase error between two data
  points separated by $100$~rotations is only \mbox{$\approx\, 1.5$~\%} so that we can neglect this as an error source here.
  
  In an effort to check how the locations of other, more distant data points compare to the lightcurves obtained during epochs~(A)
  and (B), we combine the two data sets and fit the result
  using the `free' fit approach with $9$~(independent) surface elements (Fig.~\ref{fig:HippABD}, lower panel).
  Considering the evolution between epochs~(A) and (B), the thus obtained results are of course time averages rather than
  snapshots.
  The fit provides evidence for a pronounced active region at a longitude of about $120$\textdegree \ and another
  less pronounced structure at about $240$\textdegree.

  Now we check statistically whether it is justified to assume that the time averaged model at hand remains an appropriate
  approximation of the lightcurve over a longer time span (given a period of $1.3371$~d).
  Within about one hundred days \mbox{($\approx\, 75$~rotations)} before epoch~(A) (measurements in
  \mbox{$\mathrm{JD}_H\, =\, 7950\, -\, 8010$}) there are
  $3$~other observation periods providing $8$~additional data
  points and within about the same time span after epoch~(B) (measurements in \mbox{$\mathrm{JD}_H\, =\, 8145\, -\, 8200$})
  another $7$~measurements are distributed over two observation
  epochs (see Fig.~\ref{fig:HippDataInfo}). In the middle panel of Fig.~\ref{fig:HippABD} we show the same model lightcurve as before
  as well as the $15$~additional data points (triangles and filled circles).
  %folded at the same period.
  Apparently, the model provides an acceptable
  description of the observations.
  Formally, we expect a $\chi^2$ value of \mbox{$15\, \pm\, 5.5$}
  for $15$~data points\footnote{The $\chi^2$ distribution
  with $n$~degrees of freedom has expectation value~$n$ and variance~$2n$.}
  drawn from the same model; the shown model yields $21.2$ with about
  %The number we obtain is $21.2$ with about
  one third of it contributed by a single data point.
  
  If the lightcurve evolves fast with respect to the time span under consideration,
  we expect the data points to be distributed randomly, and in the following we assume that the
  lightcurve amplitude observed in epochs~(A) and (B) remains representative.
  From a Monte Carlo simulation we then estimate that the probability to obtain
  a value of $21.2$ or less for $\chi^2$ from the same number of data points uniformly
  distributed over the lightcurve amplitude is \mbox{$\approx\, 2.5$~\%}. During the simulation we
  do not vary the data points in phase and all data points enclosed by a
  single observation epoch are regarded as dependent and, therefore, are varied as a single unit.
  
  We conclude that the Hipparcos lightcurve is compatible with a rotation period of $1.3371$~d and
  a surface configuration dominated by an active region at a longitude of about $120$\textdegree \
  for at least $50$~days and probably up to about $200$~days.
  We, however, caution the reader that the data do not allow to exclude other scenarios.
  
  The middle panel of Fig.~\ref{fig:HippABD} includes measurements carried out during the
  \mbox{$\mathrm{JD}_H\, \approx\, 8300\, -\, 8325$} time
  interval (open circles in Fig.~\ref{fig:HippABD}, also marked in Fig.~\ref{fig:HippDataInfo})
  about $210$~days after those of epoch~(B).
  The associated data points confirm the presence of a minimum
  at about the same position during this period, although their phase coverage is poor.
  %and they are probably not drawn from the model based on the epochs~(A) and (B).
  Unfortunately, phase coverage remains a problem
  throughout the analysis of the entire rest of the Hipparcos data set. Even if we combine the data obtained during
  the following \mbox{$\approx\, 450$~days} \mbox{($\mathrm{JD}_H\, =\, 8300\, -\, 8750$)}, a (connected) phase interval covering 
  about $30$~\% of the lightcurve remains unobserved (see Fig.~\ref{fig:HippDataInfo}). Furthermore, the measurements are distributed
  very inhomogeneously in time providing no further well covered individual epochs
  %so that we are forced
  forcing us to consider longer time spans.
  %for which an interpretation seems, however,
  %not unreasonable in view of the preceding discussion.
  
\begin{figure}[t]
    \includegraphics[angle=-90,width=0.5\textwidth]{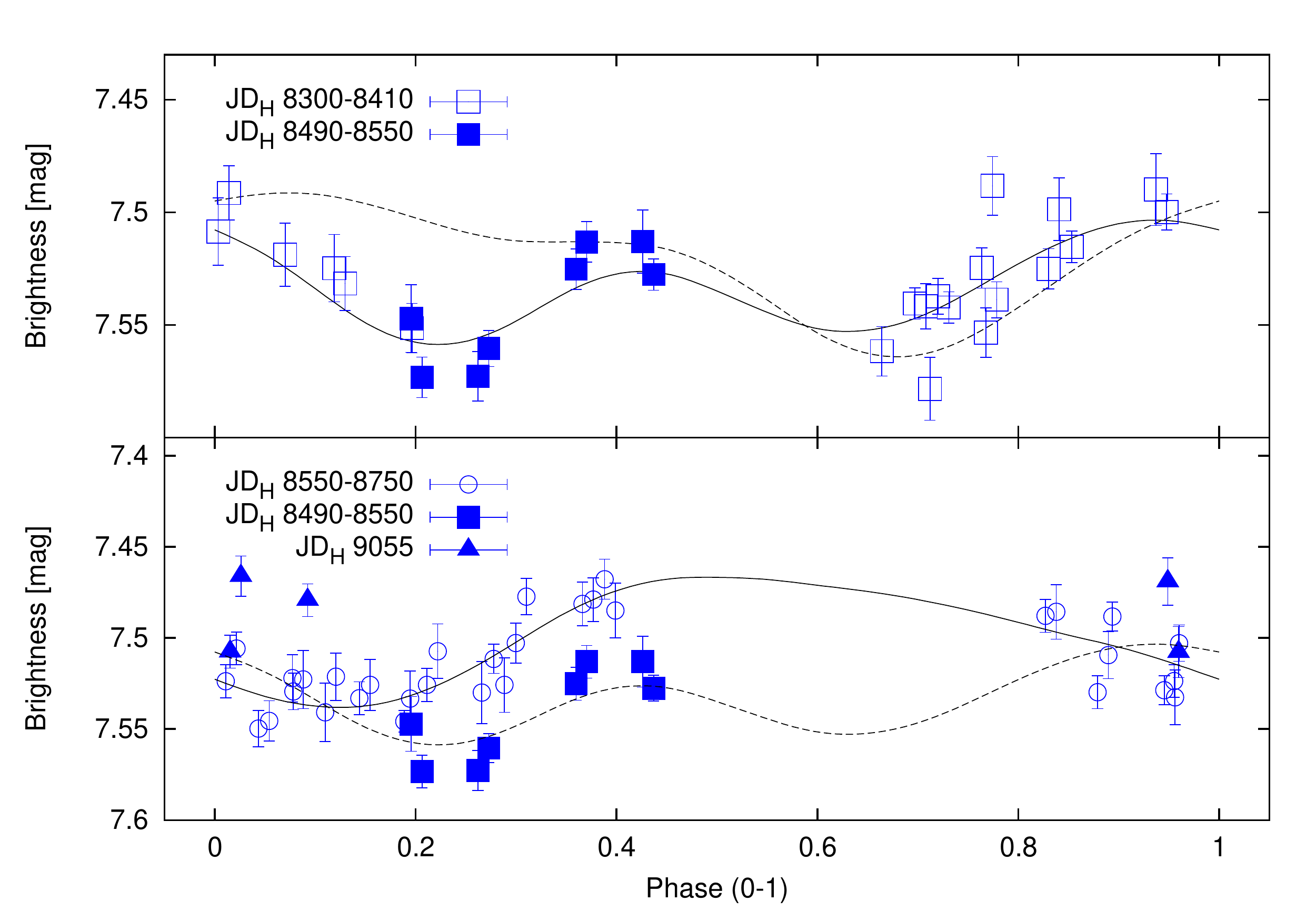}
    \caption{\textit{Upper panel:} Measurements between \mbox{$\mathrm{JD}_H\, =\, 8300$} and $8550$.
    The dashed lines indicate the model obtained from the combined epochs~(A) and (B), the solid lines show the lightcurve
    obtained from the given intervals.
    \textit{Lower panel:} Data of the \mbox{$\mathrm{JD}_H\, =\, 8490\, -\, 8750$} time interval and $5$~additional points from
    \mbox{$\mathrm{JD}_H\, =\, 9055$}.
    \label{fig:HippDE}}
  \end{figure}

  In the upper panel of Fig.~\ref{fig:HippDE} we show the data pertaining to the
  \mbox{$\mathrm{JD}_H\, =\, 8300\, -\, 8550$} time interval as well
  as a `free' model fit obtained from them; note
  that this covers some data points already shown in the middle panel of Fig.~\ref{fig:HippABD}. Additionally, we
  show the lightcurve obtained from combining epochs~(A) and (B) (dashed line).
  As already indicated earlier, both lightcurves bear considerable resemblance at phases \mbox{$\ge \, 0.5$}.
  This does no longer hold for phases \mbox{$\le \, 0.5$}, where the data clearly indicate the
  presence of a second
  pronounced minimum caused by an active region at a longitude of about $280$\textdegree.
  While this stellar region considerably increased its impact on the lightcurve compared to
  the previous data, the one at a longitude of $\approx 120$\textdegree $-$ although still existing $-$ is
  less significant and may be in a process of disintegration.
  
  During the next \mbox{$\approx\, 160$~days} (see Fig.~\ref{fig:HippDE}, lower panel), there is a clear evolution in the shape
  of the lightcurve, potentially indicating evolution of the newly emerged active region.
  Unfortunately, the phase interval sampling of the
  lower longitude active region is very poor, % covered during this observation period
  so we can no longer trace its evolution. The measurements presented in this panel
  show a trend indicating an \mbox{$\approx\, 0.05$~mag} brightening of V889~Her, which may be related to a periodic
  sinusoidal long-term trend with about the right amplitude and a period of \mbox{$\approx\, 2600$~days} detected
  by~\citet{Strassmeier2003}.
  
  Even though a stellar surface with two dominating active regions evolving in time as 
  indicated by the panels of Fig.~\ref{fig:HippDE} seems physically
  reasonable, we caution that the time span covered here is $450$~days, which is twice as long as the time span
  considered in Fig.~\ref{fig:HippABD} and, in particular, $9$~times longer than the separation between
  epochs~(A) and (B). Therefore, a variation of the rotation period in the percent-regime already has
  considerable impact on the lightcurves.
  
  Throughout the above analysis our models yield a spot coverage fraction of \mbox{$\approx\, 10$~\%} assuming a
  spot-photosphere brightness-contrast of $1/2$. Only during the last observation epochs \mbox{($\mathrm{JD}_H\, > \, 8550$)} the model provides
  a lower fraction of \mbox{$\approx\, 6$~\%}, which may, however, also be related to the poor phase coverage obtained 
  in this time span.

%%%%%%%%%%%%%%%%%%%%%%%%%%%%%%%%%%%%%%%%%%%%%%%%%%%%%%%%
\section{Radial velocity measurements}
\label{Sec:RV}

\subsection{Activity-related radial velocity measurements}
\label{Sec:RVactivity}

Stellar activity influences the strength and profiles of spectral lines and can, therefore, affect RV measurements, whose precision hinges on symmetric and narrow lines.
For example, a spot located on the hemisphere of the star rotating towards the observer diminishes the amount of light contributing to the blue-shifted line wing, leading to an asymmetric line shape.
This is not crucial for slowly rotating stars as long as their spectral lines are predominantly broadened by thermal and other mechanisms instead of rotation. Locally restricted surface features only become detectable in very broad lines where the spectra are sufficiently oversampled.
In general, slow rotators should be less active showing fewer and smaller spots, and rotational broadening is less important for their line profiles. For fast rotators with \mbox{$v \sin(i) \apprge 25$~km/s}, where all spectral lines are dominantly broadened by the Doppler effect, such surface features become resolvable in the stellar spectrum.

This is one reason why high accuracy RV measurements are difficult for rapidly rotating stars. Stellar activity can deform their line profiles and, using standard RV detection methods, these deformations lead to RV shifts due to asymmetric and variable line shapes.
RV shifts induced by stellar activity can be of the order of several 100~m/s for stars with \mbox{$v\sin(i) \, \apprge \, 5$~km/s} (\citealt{Saar1997} and \citealt{Saar1998}), and even higher for larger rotation velocities, i.e., up to a factor of thousand higher than the state-of-the-art RV precision.
While this is a nuisance for companion detections around highly active stars, it can be used as an additional source of information regarding the localization of atmospheric features.
RV modulations due to activity can be seen as an activity-induced Rossiter-McLaughlin effect \citep{Ohta2005} otherwise used for planet detection and characterization: an isolated spot on a stellar surface causes a variation of the RV curve very similar to the one caused by a transiting planet.

Nevertheless, it is a different situation with spots when using the Rossiter-McLaughlin effect.
On the one hand it is simpler, if the spot distribution does not change too rapidly, because one usually knows the rotation period of a spotted star and has a chance to observe many spot transits on small time scales for low rotation periods.
On the other hand it is more complicated because stellar spot distributions can change on short time scales and, generally, surfaces of active stars are populated by more than only one spot, which means that one measures a superposition of Rossiter-McLaughlin effects for all visible spots.
As a result, RV curves modulated by stellar activity generally do not look like the curves derived from planetary transits.
They may have complicated structures instead, which cannot be identified as obvious superpositions of Rossiter-McLaughlin effects.
However, it is possible to detect an activity-induced effect for simple configurations of spot distributions, as demonstrated in this paper.

\subsection{Rotation period}
\label{Sec:RVRP}

\begin{figure}[t!]
   \centering
   \includegraphics[width=0.5\textwidth]{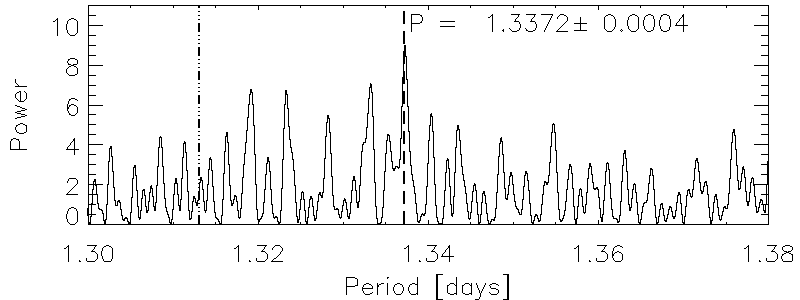}
\caption{
Periodogram of the RV data on the interval relevant for V889 Her's rotation period.
The period determined from the highest peak is given in the figure;
\citeauthor{Marsden2006}'s value of $1.313$~d is marked by the dash-dotted line.
%it deviates significantly from the value of \citet{Marsden2006}, marked by the dash-dotted line.
See Sect.~\ref{Sec:RVRP} for details.
\label{Fig:RVper}}
\end{figure}

The interpretation of long-term observations requires a precise stellar rotation period.
Data sets stretching over periods of months or years must be folded back at a well-known phase scale to be able to detect any long-term stability.
In the case of V889 Her this is possible on the basis of extensive photometric observations yielding a rotation period of \mbox{$P \, = \, 1.3371 \, \pm \, 0.0002$~days} \citep{Strassmeier2003}.
Note that we obtain the same value from a periodogram of our RV measurements, presented in Fig.~\ref{Fig:RVper}, where the most significant peak in the relevant period interval is located at \mbox{$P \, = \, 1.3372 \, \pm \, 0.0004$~days}.

The position of \citeauthor{Marsden2006}'s rotation period is marked as well
in Fig.~\ref{Fig:RVper}; clearly, our RV data support the period determined by \citeauthor{Strassmeier2003}, which we use in the following.
This rotation period applies to the case of rigid rotation; since no significant evidence for differential rotation was found in our data sets, we hold on to the simple model (see Sect.~\ref{Sec:IntDis}).
%As discussed in Sec.~\ref{Sec:IntDis}, our observations are only compatible with differential rotation if the dominant features are confined to $\sim\, 30$\textdegree \ latitude.}
%\textbf{Note that this is not the only possible rotation period if differential rotation is expected \citep{Marsden2006}; with our data sets we cannot exclude this possibility although we argue in favor of small differential rotation (see Sect.~TBD).}

Given \citeauthor{Strassmeier2003}'s error of the rotation period \mbox{$\Delta P \, = \, 0.0002$~days} and considering an acceptable phase error of \mbox{$\Delta \phi \, = \, 0.05$} (translating into \mbox{$\approx \, 20$\textdegree}), we can calculate the time span \mbox{$\Delta \mathrm{JD}$} that can be phase-folded with the required accuracy.
% The selected maximum phase error is associated with the longitude accuracies of \mbox{$\apprle \, \pm \, 10$\textdegree} achieved in our 2-spot models (cf. Sect.~\ref{Sec:StableHIPP}).
The selected maximum phase error is consistent with longitude accuracies of \mbox{$\apprle \, \pm \, 10$\textdegree} achieved during the modeling of large active regions in the Hipparcos photometry.
We use equation
\begin{equation}
\Delta \mathrm{JD} \ = \ \left| \left( \frac{1}{P + \Delta P} - \frac{1}{P} \right)^{-1} \cdot \Delta \phi \right| \ ,
\label{Eq:DeltaJD}
\end{equation}
derived from \mbox{$\phi + \Delta \phi \, = \, \Delta \mathrm{JD}/(P + \Delta P) $}, %(Eq.~\ref{Eq:phases})
to determine a maximum duration of \mbox{$\Delta \mathrm{JD} \ \approx \ 450$~days} for which we can trace a structure in the lightcurve with a phase accuracy better than $0.05$.

We use \mbox{$\mathrm{JD}_0 \, = \, 2449000.0$} to calculate the phases $\phi$ from
\begin{equation}
\phi = \frac{\mathrm{JD} \ - \ \mathrm{JD}_0}{P_{rot}} \ .
\label{Eq:phases}
\end{equation}
Stellar surface coordinates are defined consistently for all surface maps (see e.g. Fig.~\ref{Fig:DIimages}).
Rotation phases are calculated for surfaces rotating towards decreasing longitudes, which means that a phase $\phi$ can be translated into a longitude $l$ using the equation \mbox{$\phi \ = \ (360^{\circ}-l)/360^{\circ}$}.

\begin{figure*}[t!]
  \centering
    \includegraphics[width=0.32\textwidth]{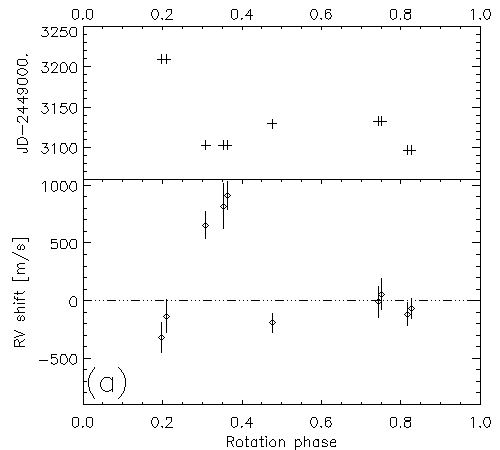}
    \includegraphics[width=0.32\textwidth]{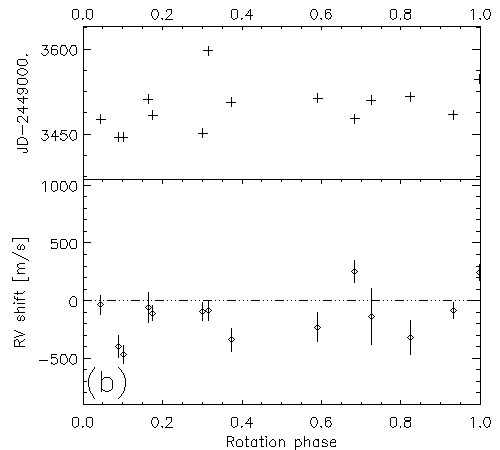}
    \includegraphics[width=0.32\textwidth]{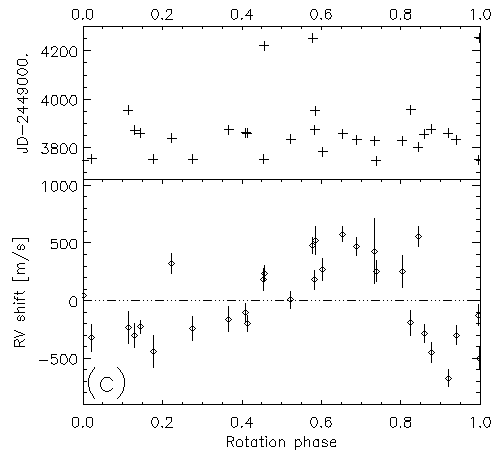} \\
    \includegraphics[width=0.32\textwidth,clip=,bb=0 20 500 450]{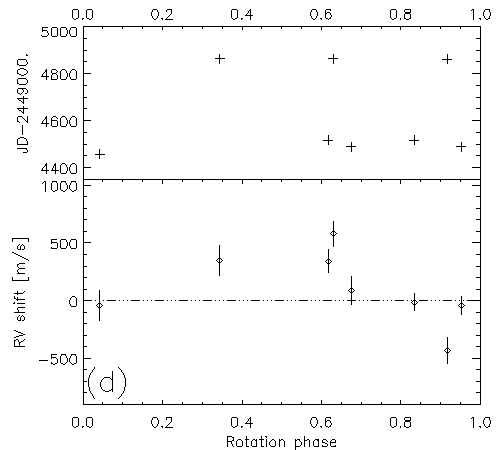}
    \includegraphics[width=0.32\textwidth,clip=,bb=0 20 500 450]{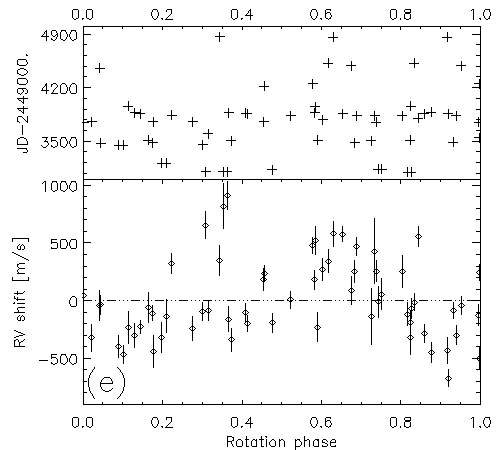}
    \includegraphics[width=0.32\textwidth,clip=,bb=0 20 500 450]{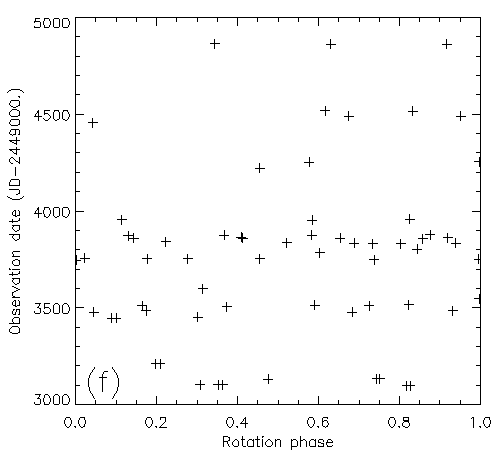}
    \vskip6pt \hrule \vskip5pt
    \includegraphics[width=1.\textwidth]{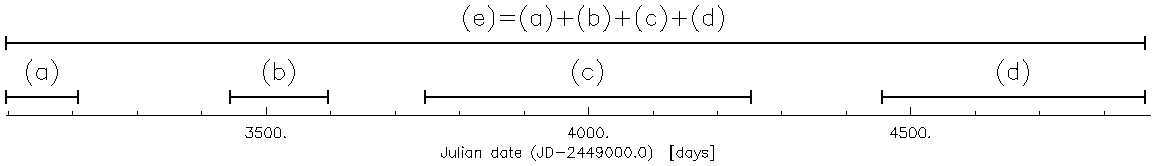}
    \vskip2pt \hrule
  \caption{Radial velocity measurements of V889 Her.
The data set was separated into several parts to visualize the stability of structures in the RV curve.
Each interval of a few hundred days shows characteristic influences of the star's activity on the spectra causing RV shifts.
Large spotted regions are visible in the RV curves for more than a hundred days before they finally dissolve or significantly change positions.
Panel~(f) shows phase coverage of the entire data set.
\label{Fig:RVdata}}
\end{figure*}

\subsection{Detection of stable active regions in the RV curves}
\label{Sec:RVstable-regions}

The RV measurements of V889 Her cover a time period of almost $2\,000$~days, which translates into approximately $1\,500$~stellar rotations.
The distribution of data points over the entire observations is given in Fig.~\ref{Fig:RVdata} panel~(f);
in panels~(a) to (e) we subdivide the data into shorter intervals.
%(see Sects.~\ref{Sec:LCsep} and~\ref{Sec:RVRP} for a detailed discussion on the necessity of such subintervals).
%\textbf{where the RV pattern appears to provied higher stability.}
The lower panel illustrates the time coverage of each panel.

The modulation of the RV curve does not reveal any stellar companion or extrasolar planet within the limits of its accuracy. There are large quasi-periodic variations in the curve changing on large time scales of several months, which is not consistent with a low-mass companion. The average accuracy per data point is about $150$~m/s.

Nevertheless, we do detect periodicity in the RV curve of Fig.~\ref{Fig:RVdata}.
The periodogram of the data shows a high peak at the star's rotation period (cf. Sect.~\ref{Sec:RVRP}), which suggests that the modulation of the curve is related to its activity.
In Fig.~\ref{Fig:RVdata} subintervals of the RV data are presented, where the global shape of the curve does not change significantly;
however, between the panels substantial changes of the RV curve take place.
Panel~(c) contains a large fraction of all data points (almost 50~\%) and clearly shows a stable modulation of the RV curve with rotation for a duration of at least \mbox{$\apprge \, 150$~days} \mbox{($\mathrm{JD}\, = \, 3750\, -\, 3900$)};
this time span may even be longer since neighboring points from \mbox{$\mathrm{JD}\, \approx\, 3950$} and $\approx\, 4250$ support the curve as well.

% As discussed in Sect.~\ref{Sec:DIrvlc}, this curve agrees well with Doppler images obtained from data inside the same time interval.
% The modulation is due to an activity-induced Rossiter-McLaughlin effect caused by the stellar spot distribution and
% confirms that the global spot distribution of V889 Her was stable within at least one year.
% To our knowledge, this is the first time that activity-related RV measurements are successfully compared to reconstructed spot distributions,
% which shows that RV variations can be a powerful tool for stellar activity research as well.

The RV modulation visible in panel (c) dominates the RV curve over the entire time interval, as visible in panel~(e) where the phase-folded RV curve of all data points is presented. This is the reason why the periodogram returns the star's exact rotation period, although the spot distribution changes on long time scales. It may not be possible to obtain the rotation period from periodograms of RV curves for a faster changing spot distribution. Its effects on the RV curve will shift in phase and change in amplitude, which is not compensated by regular periodogram algorithms. If the rotation period is known and a sufficiently high phase coverage is available, the data can be subdivided into intervals with stable RV modulations associated to stable global spot distributions.

It is difficult to detect activity-induced Rossiter-McLaughlin effects (see Sect.~\ref{Sec:RVactivity}) in the other panels of Fig.~\ref{Fig:RVdata} due to the poorer data sampling.
Panel~(a) possibly contains some evidence for such a structure but further complementary data, i.e., Doppler images or lightcurves, are necessary to confirm this assumption.
Panel~(b) does not show any modulation implying that the spot distribution of panel~(c) must have formed (or strongly intensified) after \mbox{$\mathrm{JD} \, \approx \, 3500$}.
There are possibly some signatures of this modulation left in panel~(d); the rather poor phase coverage makes it hard to determine to which degree the spot distribution changed.

%%%%%%%%%%%%%%%%%%%%%%%%%%%%%%%%%%%%%%%%%%%%%%%%%%%%%%%%
\section{Doppler Imaging}
\label{Sec:DI}

Surface inhomogeneities cause deformations of spectral line profiles which can be used to reconstruct the 
spot distribution. This is achieved by Doppler Imaging \citep{Vogt1983}.
DI is limited to fast rotating stars where the broadening of the spectral lines is dominated by %the Doppler effect.
rotational broadening.
Given the rotation velocity and period of V889~Her, surface reconstructions of this star can be achieved
%(cf.\citet{Strassmeier2003,Marsden2006}).
(cf.~\citeauthor{Strassmeier2003}, 2003; \citeauthor{Marsden2006}, 2006).

\subsection{CLEAN-like Doppler Imaging algorithm}
\label{Sec:DIclean}

For Doppler Imaging (DI) we use the CLEAN-like algorithm `CLDI' \citep{WolterPHD, Wolter2005a}.
%Like any DI algorithm CLDI requires a time series of line profiles with good phase coverage.
Starting with a `standard' line profile adopted for the unspotted star, 
the program deforms the line profiles using an iteratively constructed stellar surface in 
order to match the shape of the spectral lines for all observed phases.
This is done using `probability maps' that show regions of tentative spot locations \citep{Kuerster1993}.
These maps do not give a probability in a strict statistical sense, instead they are `backprojections' 
of the derivations between the observed and reconstructed line profiles onto the stellar surface.
Spots are set at the most intense surface element of these backprojections as long as the
deviations between the observed and reconstructed line profiles keep decreasing.

Several parameters must be supplied a priori to the DI process:
the rotation period $P$, the projected rotation velocity \mbox{$v \sin(i)$},
the stellar inclination $i$, and the linear limb darkening parameter $\epsilon$
%We use a linear limb darkening law  \citep{GrayBOOK},
%i.e. 
(with \mbox{$I_C / I^0_C \, = \, 1 - \epsilon + \epsilon \cos \Theta$}
where \mbox{$\cos \Theta$} is the projected distance
from the center of the stellar disk in units of the stellar radius \citep{GrayBOOK}).
%A list of these parameters for all Doppler images of this paper can be found in 
Table~\ref{Tab:DIpar} contains the values of these parameters adopted for our DI,
additional stellar properties are given in Table~\ref{Tab:V889prop}, while 
Tab.~\ref{Tab:DIlines} lists the lines used for the surface reconstructions.

The Doppler maps are reconstructed in terms of a spot filling factor between
 $0$ (no spot) and $1$ (completely spotted) where
% for each surface element; 
the number of possible graduations $n_T \, \ge \, 2$ must be preselected.
Spotted surface elements are areas with lower continuum intensity than the unspotted stellar surface, 
CLDI does not explicitly determine spot temperatures.
%The stellar inclination \mbox{$i \, = \, 55$\textdegree} was adopted from \citet{Strassmeier2003}.
We find no differential rotation within our detection limits;
reliable small-scale structures close to the pole are hardly resolved 
preventing the determination of a polar rotation period (see Fig.~\ref{Fig:DIimages}).
Reconstructions of the star taken several rotation periods apart do not yield significant
identifications of systematically shifted features (see Appendix Fig.~\ref{Fig:DIcrosscor} for cross-correlation maps).
Doppler images calculated using \citeauthor{Marsden2006}'s differential rotation law
do not improve $\chi^2$.
%Reconstructions of different rotations do not yield significant identifications of systematically
%shifted features.
Therefore, we refrain from modifying the rigid rotation law used in our reconstructions.
The limb darkening parameter was adjusted for each line to obtain an optimal fit of the 
standard line profile to the observed lines, resulting in values between $\epsilon \ = \ 0.6$ 
and $0.8$.

Based on a comparison of our reconstructions from different 
spectral lines (Fig.~\ref{Fig:DIimages}),
%the noise and spectral resolution, 
we estimate their surface resolution
as approximately $10$\textdegree \ corresponding to \mbox{$2 \times 2$} 
surface elements;
a few features are poorer localized in latitude (e.g. feature~C in the left column maps
of Fig.~\ref{Fig:DIimages}). 
Isolated surface elements are not reliably localized,
they are an artifact of the CLDI algorithm due to noise 
and/or short-term spot evolution on small spatial scales.

The estimated reduced $\chi^2$ values of our reconstructions, averaged over all phases, are between $0.3$ and $1.5$.
It is not possible to provide a strictly reduced $\chi^2$, since the number of parameters is not known in the
CLEAN-like approach; however, the values are close to $1$ and the reconstruction show good agreement
with the line profiles (see Appendix~\ref{Fig:DI-lineprofiles}).

\subsection{Doppler Imaging of V889 Her}
%\label{Sec:DIresults}

\begin{figure*}[t!]
  \begin{center}
    \begin{tabular}{c c || c}
      \hline \hline
      & Nights \textbf{1 - 4} & Nights \textbf{9 - 12} \\
      \hline
      \rotatebox{90}{\hspace*{1.5cm} \textbf{6122} \AA \hspace*{1.5cm} } \hspace*{-.3cm} &
      \includegraphics[width=0.45\textwidth]{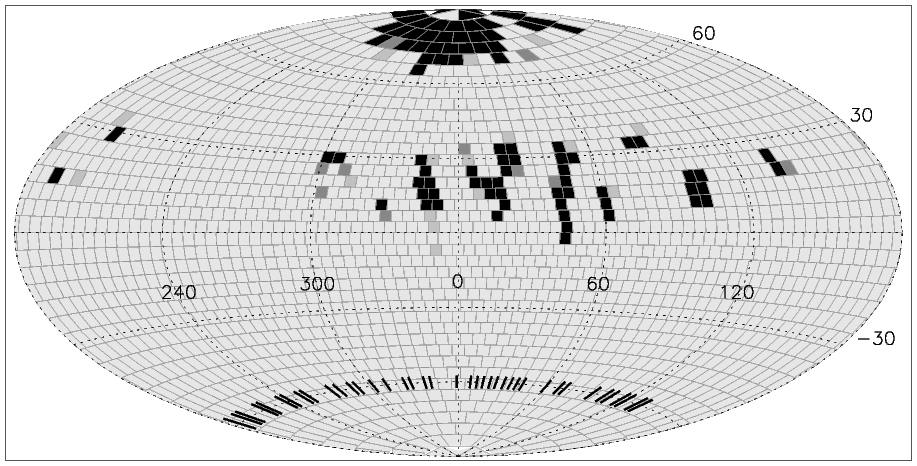} &
      \includegraphics[width=0.45\textwidth]{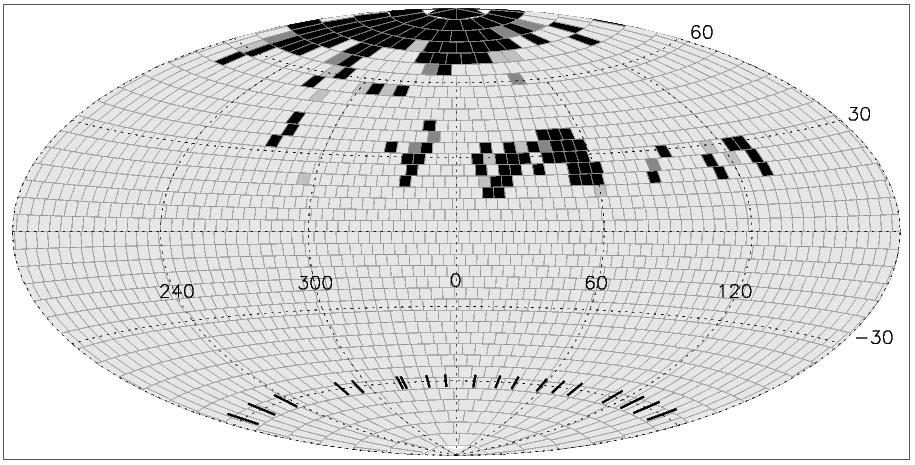} \\
      \hline
%       \rotatebox{90}{\hspace*{1.5cm} \textbf{6265} \AA \hspace*{1.8cm}} \hspace*{-.3cm} &
%       \includegraphics[width=0.45\textwidth]{pics/DI-6265-14eqw-hig132_cut.png} &
%       \includegraphics[width=0.45\textwidth]{pics/DI-6265-58eqw-hig132_cut.png} \\
%       \hline
      \rotatebox{90}{\hspace*{1.5cm} \textbf{6439} \AA \hspace*{1.6cm}} \hspace*{-.3cm} &
      \includegraphics[width=0.45\textwidth]{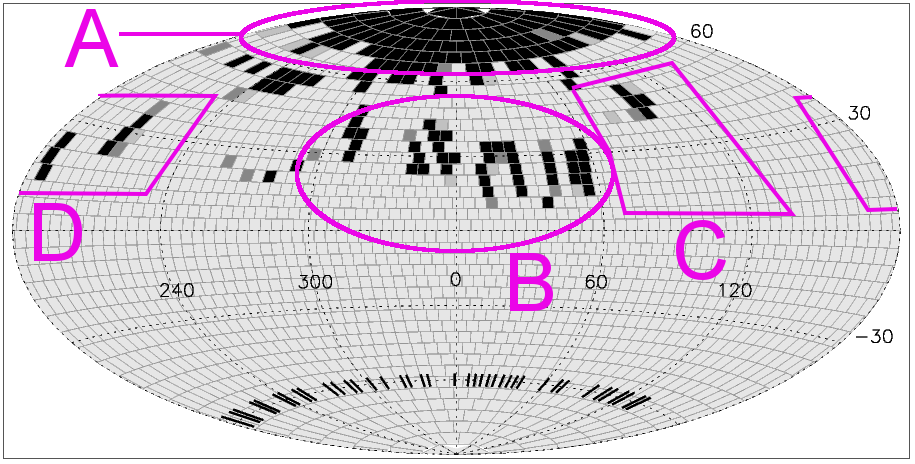} &
      \includegraphics[width=0.45\textwidth]{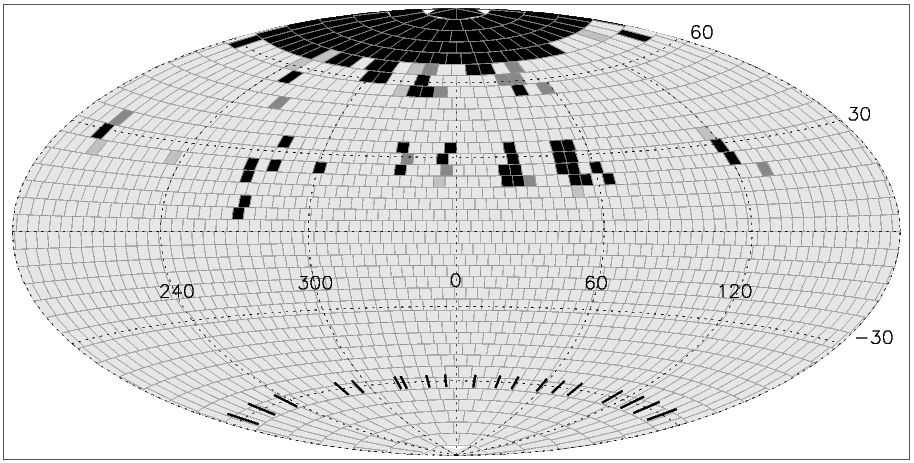} \\
      \hline
%       \rotatebox{90}{\hspace*{1.5cm} \textbf{6462} \AA \hspace*{1.8cm}} \hspace*{-.3cm} &
%       \includegraphics[width=0.45\textwidth]{pics/DI-6462-14eqw-reg_cut.png} &
%       \includegraphics[width=0.45\textwidth]{pics/DI-6462-58eqw-reg_cut.png} \\
%       \hline
    \end{tabular}
  \end{center}
\caption{Doppler images of V889 Her, each observed during three stellar rotations.
The images in the left column are separated from the right ones by three rotations.
Each row contains images reconstructed from the indicated spectral line.
Ticks in the southern hemispheres indicate the observed phases used for the reconstruction.
Letters~\mbox{A$-$D} annotate spot regions that can be commonly identified in several reconstructions.
See Sects.~\ref{Sec:DIclean} and~\ref{Sec:DIimages} for discussion.
\label{Fig:DIimages}}
\end{figure*}

%\subsubsection{Surface reconstructions}
\label{Sec:DIimages}

\begin{table}[b!]
  \begin{minipage}[h]{0.5\textwidth}
  \renewcommand{\footnoterule}{}
  \caption{Absorption lines used for Doppler Imaging.
    \label{Tab:DIlines}}
\begin{center}
    \begin{tabular}{l c c}
      \hline \hline
      Name \hspace*{1cm} & Rest wavelength\footnote{from VALD (Vienna Atomic Line Database)} (\AA) & Chemical element \\
      \hline
      6122  & 6122.22 & Ca I \\
%      6265  & 6265.14 & Fe I \\
      6439  & 6439.08 & Ca I \\
%      6462  & 6462.57 + 6462.73 & Ca I + Fe I \\
      \hline
    \end{tabular}
\end{center}
  \end{minipage}
\end{table}

Figure~\ref{Fig:DIimages} shows four reconstructed Doppler images of V889~Her.
The left column contains the images of the first full rotation, 
based on spectra of four consecutive nights (`nights~$1\, -\, 4$'), i.e. three stellar rotations.
The right column shows images of the second full rotation (`nights~$9\, -\, 12$'), also
taken during three consecutive stellar rotations.
Three stellar rotations remained unobserved between the images of the two columns.
Table~\ref{Tab:DIlines} contains details of the lines used for DI.

The left column maps are based on 39 observed rotation phases, quite evenly distributed apart from a gap
ranging from about 130\textdegree \ to 180\textdegree \ in surface longitude,
while the right column maps are based on 19 phases.
The left 6439~\AA-map labels the spot regions A$\, -\,$D, that can be commonly identified in several reconstructions.
In agreement with previous reconstructions of V889~Her, all our maps show a large polar spot.
Causing almost no rotational modulation, the reconstructed size of a polar spot depends sensitively
on the adopted line parameters.
However, keeping these uncertainties in mind, two properties of the polar spot are
reliably found in our reconstructions: it extends down to a latitude of approximately +60\textdegree \
and it apparently exhibits a weak asymmetry, being slightly more pronounced on the 0$\, -\,$180\textdegree \
hemisphere (right half of each map in Fig.~\ref{Fig:DIimages}). 

\begin{table}[b!]
  \begin{minipage}[h]{0.5\textwidth}
  \renewcommand{\footnoterule}{}
  \caption{Reconstruction parameters adopted for DI.
    \label{Tab:DIpar}}
\begin{center}
    \begin{tabular}{l c}
      \hline \hline
      Parameter \hspace*{3.5cm} & Value \\
      \hline
      Differential rotation $\alpha$                     & 0. \\
      Limb darkening $\epsilon$                          & $0.6\, -\, 0.8$ \\
%      Inclination $i$ \footnote{\citet{Strassmeier2003}} & 55\textdegree \\
      Macro-turbulence \footnote{Probably unphysical since mainly used for line profile adjustments.} % Low value for 6265~\AA \ line, high value for all other lines.}
                                                         & 2 / 12 $\mathrm{km} \, \mathrm{s}^{-1}$ \\
      Contrast photosphere/spot                          & 0.5 \\
      $n_T$ \footnote{number of intensity levels between darkest and brightest surface elements (`surface temperatures')}
                                                         & 4 \\
      \hline
    \end{tabular}
\end{center}
  \end{minipage}
\end{table}

% \begin{figure*}[t!]
%   \begin{center}
%    \parbox{0.24\textwidth}{Phase 0.8 (Interval 1, end)} \parbox{0.24\textwidth}{Phase 0.95 (Interval 2, center)}
%    \parbox{0.24\textwidth}{Phase 0.1 (Interval 3, start)} \parbox{0.24\textwidth}{Phase 0.45 (Interval 4, center)} \\
%    \includegraphics[width=0.24\textwidth]{pics/DI-6439-14eqw-reg-map-phase08cl_new.png}
%    \includegraphics[width=0.24\textwidth]{pics/DI-6439-14eqw-reg-map-phase095cl_new.png}
%    \includegraphics[width=0.24\textwidth]{pics/DI-6439-14eqw-reg-map-phase01cl_new.png}
%    \includegraphics[width=0.24\textwidth]{pics/DI-6439-14eqw-reg-map-phase045cl_new.png}
%   \end{center}
% \caption{
% Doppler image of V889 Her (nights~$1\, -\, 4$, 6439~\AA \ line, see Fig.~\ref{Fig:DIimages})
% shown at selected phases that serve the discussion of RV curves of Fig.~\ref{Fig:DIRVLC}.
% The phase intervals~1 to~4 annotated on top of each panel are defined there.
% The sequence illustrates the movement of the dominant spot group (region B) across the visible disk:
% full visibility at the left limb (1), centered on the stellar disk (2), and movement to the right limb (3).
% In panel~(4) the spot group is located at the back side of the star.
% Qualitatively, the same sequence can be followed in all our Doppler images.
% \label{Fig:DI6439-phases}}
% \end{figure*}

Both the dominant low-latitude feature B and its much smaller companion C persist through all 9~rotations
covered by our Doppler maps. They show little distinct evolution apart from a slightly more pronounced
gap in the middle of feature~B during nights~$9\,-\,12$. 
The other low-latitude feature~(D) dissolves between the two observed sequences.

When interpreting Doppler images, one must keep in mind that spots located on the southern hemisphere
are largely projected onto the northern hemisphere. Therefore, all given spot latitudes have an unspecified sign.
Additionally, surface features based on poor phase coverage have a higher uncertainty and must be judged with care.
The latter aspect does not apply to any of the features discussed above.

\begin{figure*}[t!]
  \begin{center}
    \begin{tabular}{c c c}
%       \hline \hline
%       & Measured RV shift of V889 Her & Legend \\
%       \hline
      \rotatebox{90}{\hspace*{0.8cm} Measured RV curve}
      & \includegraphics[width=0.40\textwidth]{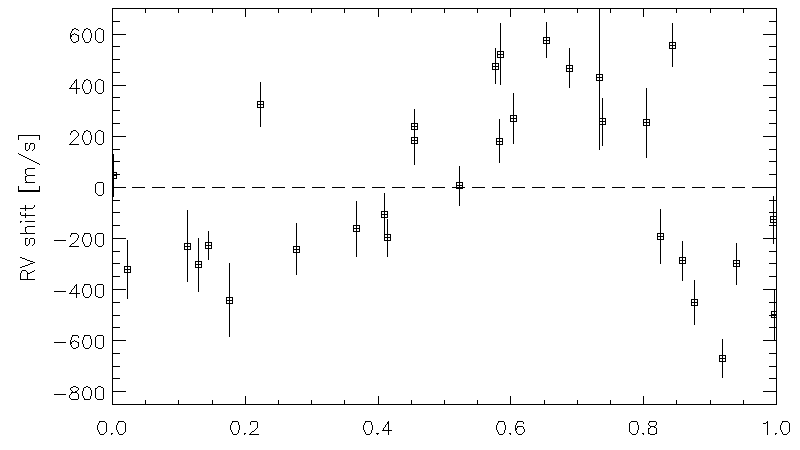}
      & \includegraphics[width=0.40\textwidth,clip=,bb=0 0 800 450]{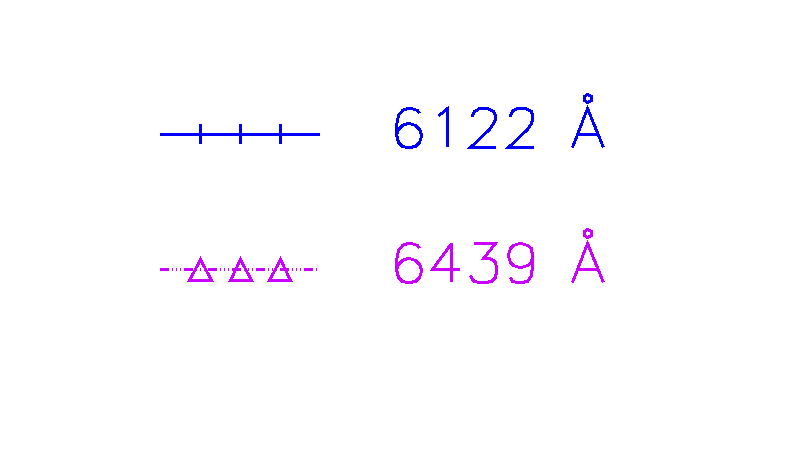} \\
%       \hline \hline
%       & \hspace*{1cm} Line barycenters for the nights \textbf{1 - 4} \hspace*{1cm} 
%       & \hspace*{1cm} Light curves for the nights \textbf{1 - 4} \hspace*{1cm} \\
%       \hline \vspace*{-2pt}
%       \rotatebox{90}{\hspace*{0.5cm} \textbf{6122} \AA \ \ \& \ \ \textbf{6265} \AA \hspace*{.1cm} } \hspace*{-.5cm} &
      \rotatebox{90}{\hspace*{0.8cm} Nights 1 $-$ 4} &
      \includegraphics[width=0.40\textwidth,clip=,bb=0 35 800 460]{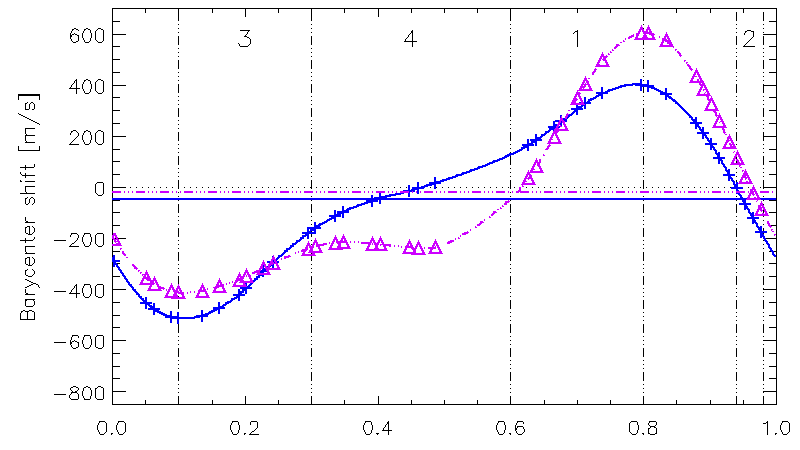} &
      \includegraphics[width=0.40\textwidth,clip=,bb=0 35 800 460]{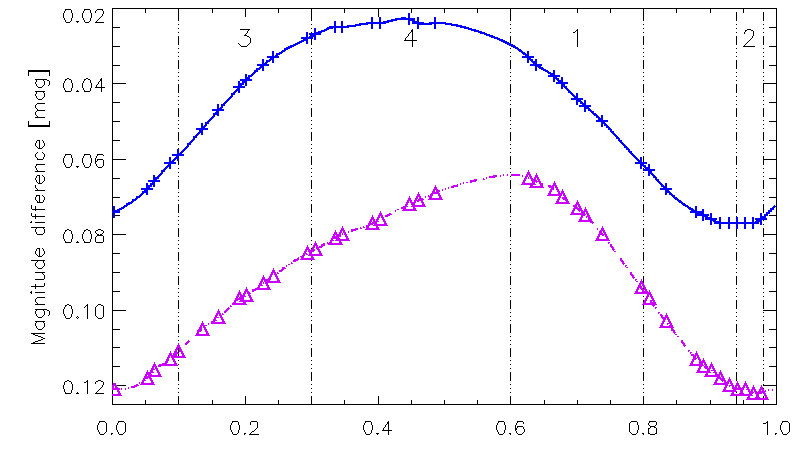} \\
%       \rotatebox{90}{\hspace*{.5cm} \textbf{6439} \AA \ \ \& \ \ \textbf{6462} \AA \hspace*{.1cm} } \hspace*{-.5cm} &
%       \includegraphics[width=0.39\textwidth,clip=,bb=0 0 800 440]{pics/DI-centershift14-line6439a6462.png} &
%       \includegraphics[width=0.39\textwidth,clip=,bb=0 0 800 440]{pics/DI-lightcurve14-line6439a6462.png} \\
%       \hline
%       & \hspace*{1cm} Line barycenters for the nights \textbf{5 - 8} \hspace*{1cm} 
%       & \hspace*{1cm} Light curves for the nights \textbf{5 - 8} \hspace*{1cm} \\
%       \hline \vspace*{-2pt}
%       \rotatebox{90}{\hspace*{0.5cm} \textbf{6122} \AA \ \ \& \ \ \textbf{6265} \AA \hspace*{.1cm} } \hspace*{-.5cm} &
      \rotatebox{90}{\hspace*{1.2cm} Nights 9 $-$ 12} &
      \includegraphics[width=0.40\textwidth,clip=,bb=0 0 800 440]{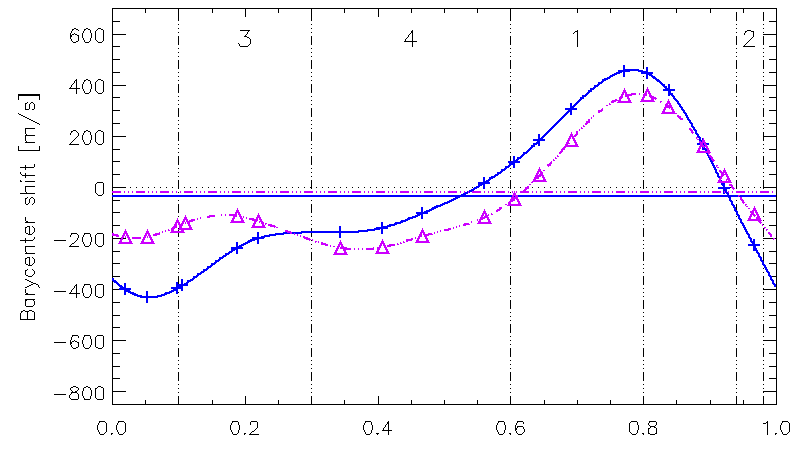} &
      \includegraphics[width=0.40\textwidth,clip=,bb=0 0 800 440]{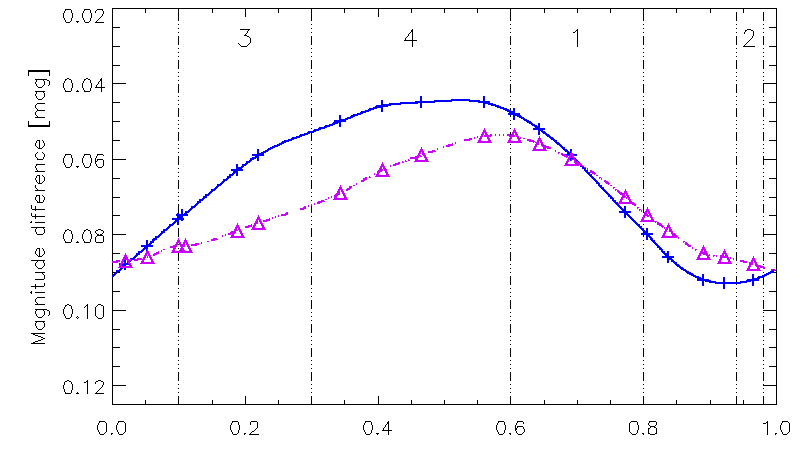} \\
%       \rotatebox{90}{\hspace*{.5cm} \textbf{6439} \AA \ \ \& \ \ \textbf{6462} \AA \hspace*{.1cm} } \hspace*{-.5cm} &
%       \includegraphics[width=0.39\textwidth,clip=,bb=0 0 800 440]{pics/DI-centershift58-line6439a6462.png} &
%       \includegraphics[width=0.39\textwidth,clip=,bb=0 0 800 440]{pics/DI-lightcurve58-line6439a6462.png} \\
%       & \hspace*{.8cm} {\small Rotation period} & \hspace*{.8cm} {\small Rotation period} \\
%       \hline
    \end{tabular}
  \end{center}
\vspace*{-12pt}
\caption{
Line barycenters and lightcurves of V889 Her around $\mathrm{JD}_0 \, + \, 4 \, 000$.
On top the observed RV curve is presented (interval~(c) of Fig.~\ref{Fig:RVdata}).
\textit{Left column:}
Line barycenters computed from our DI line profile reconstructions (see Sect.~\ref{Sec:DIrvlc});
they all show an activity-induced Rossiter-McLaughlin effect matching the observations well.
\textit{Right column:}
Lightcurves computed from our Doppler images showing a characteristic correlation to the barycenter modulations.
The defined phase intervals~$1\, -\, 4$ are used in the discussion of Sect.~\ref{Sec:DIrvlc-res}, see also Fig.~\ref{Fig:DI6439-phases}.
\label{Fig:DIRVLC}}
\end{figure*}

%We conclude that the global spot distribution of V889 Her hardly changed on the time scale of a few rotations,
%covered by our DI~data.
%% UW 2009feb
%\textbf{We conclude that
%the spot distribution of V889~Her does not exhibit pronounced changes on the time scale of a few rotations
%covered by our DI~data.
%the quality of our DI~data do not allow us to analyze spots on smaller scales than discussed above.
We conclude that all maps show large spotted regions on poles and low latitudes.
Although these features \mbox{(A $-$ D)} seem to be subject to evolution,
their change in fine structure is significantly affected by reconstruction uncertainties.
However, some spot groups are similar in their large scale structure, especially region~B
which keeps its character as `dominant feature' during the observations;
this is also visible in the lightcurves and RV-curves of Fig.~\ref{Fig:DIRVLC}.
Within our accuracy limits we do not find evidence of differential rotation:
the identified low-latitude surface features do not show a systematic shift in longitude 
during the observed 9~rotations.
%neither does the polar spot exhibit well-defined asymmetries
%which would be required to measure a change in rotation period for higher latitudes.
%% UW 2009feb
Neither does the polar spot or its lower-latitude appendices exhibit well-defined asymmetries
which would be required to measure a differing rotation period for higher latitudes.
This justifies our choice of a rigid rotation for the surface reconstruction in agreement with \citep{Strassmeier2003}.
To obtain significant measurements of V889~Her's differential rotation,
considerably less noisy line profiles would be required as input of the DI.

\begin{figure*}
  \begin{center}
   \parbox{0.24\textwidth}{Phase 0.8 (Interval 1, end)} \parbox{0.24\textwidth}{Phase 0.95 (Interval 2, center)}
   \parbox{0.24\textwidth}{Phase 0.1 (Interval 3, start)} \parbox{0.24\textwidth}{Phase 0.45 (Interval 4, center)} \\
   \includegraphics[width=0.24\textwidth]{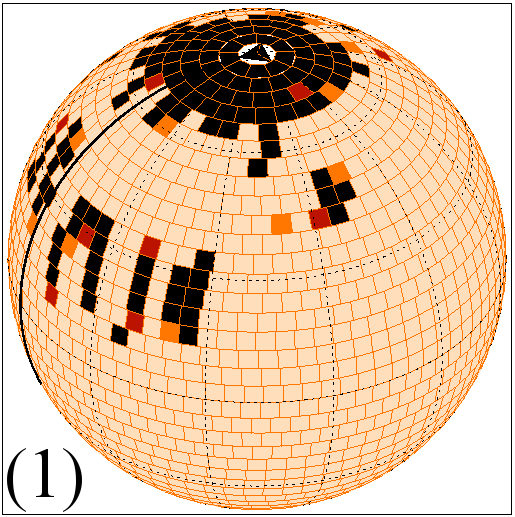}
   \includegraphics[width=0.24\textwidth]{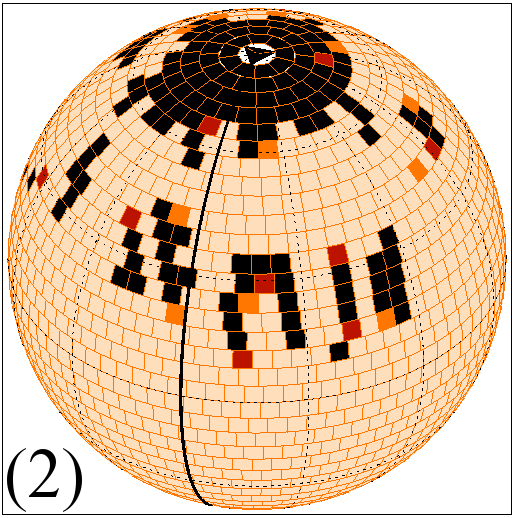}
   \includegraphics[width=0.24\textwidth]{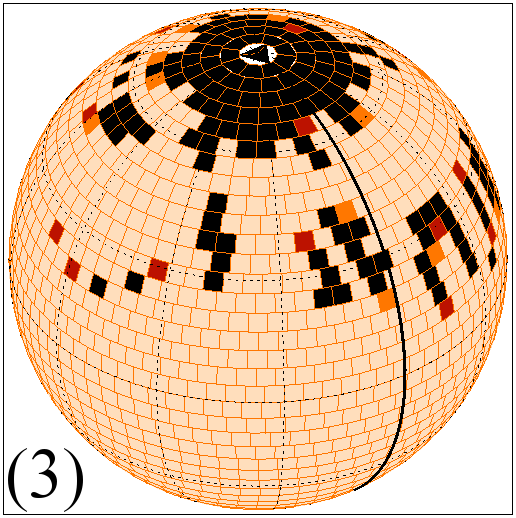}
   \includegraphics[width=0.24\textwidth]{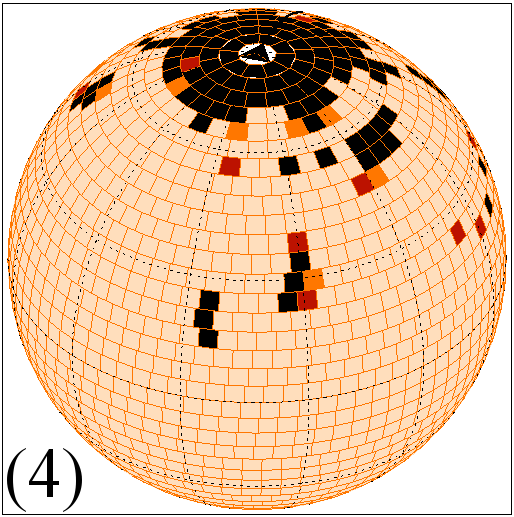}
  \end{center}
\caption{
Doppler image of V889 Her (nights~$1\, -\, 4$, 6439~\AA \ line, see Fig.~\ref{Fig:DIimages})
shown at selected phases that serve the discussion of RV curves of Fig.~\ref{Fig:DIRVLC}.
The phase intervals~1 to~4 annotated on top of each panel are defined there.
The sequence illustrates the movement of the dominant spot group (region B) across the visible disk:
full visibility at the left limb (1), centered on the stellar disk (2), and movement to the right limb (3).
In panel~(4) the spot group is located at the back side of the star.
Qualitatively, the same sequence can be followed in all our Doppler images.
\label{Fig:DI6439-phases}}
\end{figure*}

\subsection{RV shifts and lightcurves obtained by Doppler Imaging}
\label{Sec:DIrvlc}

For our Doppler images we found ourselves in the unusual situation that we
%lack contemporaneous photometry, but
had contemporaneous precise and time-resolved RV measurements.
%instead.
In the following we describe how these can be used to obtain activity information and
how they can be compared to Doppler maps.
%to check the consistency of the latter with our Doppler maps.

\subsubsection{Approach}

Asymmetric deformations of a line profile lead to an apparent radial velocity shift of the line.
%The line profile of an unspotted star is symmetric and constant for all rotation phases,
%at least for the resolution and data quality of typical DI spectra.
%We follow the idea that variable asymmetries in line profiles reflect spots moving across the stellar disk.
To quantify such profile asymmetries, we define the center of a line profile as its `barycenter':
if $x$ is the radial velocity measured in units of \mbox{$v \sin(i)$} and $y(x)$ is the corresponding (normalized) flux,
we can write down the equation
\begin{equation}
S \ = \
- \sum \limits_{i=1}^N \frac{x_i \ y(x_i) \ \Delta x}{W_{\mathrm{Eq}}} \ = \
- \frac{2 \ x_{\mathrm{max}}}{N \ W_{\mathrm{Eq}}} \sum \limits_{i=1}^N x_i \ y(x_i) \ .
\label{Eq:bary}
\end{equation}
$S$ denotes the barycenter of a line profile in analogy to the barycenter of a system of masses,
$W_\mathrm{Eq}$ is the spectral line's equivalent width, and $\Delta x$ is the interval between $x_i$ and $x_{i+1}$.
Applying Eq.~\ref{Eq:bary} to our reconstructed line profiles computed from our DI, 
we obtain a measure for the asymmetry-induced shift of the line center 
as a function of rotation phase.
Note that this is not necessarily the same value as derived in standard RV 
measurements. Nevertheless, as shown in Fig.~\ref{Fig:DIRVLC}, in our case both approaches yield
similar RV curves for the same object. This similarity encompasses amplitude and shape of the
RV curves.
Eq.~\ref{Eq:bary} is discussed in detail in Appendix~\ref{App:Eq1}, see also~\citet{Ohta2005}.

%Lacking contemporaneous photometry, we could not use light curves to verify the large-scale spot distribution
%of our Doppler images.
Although we do not have contemporaneous photometry, we compute lightcurves from our images mainly for two reasons.
%for other purposes.
First, as Fig.~\ref{Fig:DIRVLC} illustrates, they facilitate the interpretation of the RV curves.
Second, they can be compared \textit{qualitatively} to the Hipparcos lightcurves
of Figs.~\ref{fig:HippABD} and~\ref{fig:HippDE}, regarding both amplitude and shape.
Using the pre-defined spot continuum flux and limb darkening law, it is straight forward to compute a lightcurve from our Doppler maps.
For each rotation phase a disk-integrated flux $f_i$ is calculated and transformed to 
magnitudes using \mbox{$m \ = \ - 2.5 \log (f_i / f_0) $} with \mbox{$f_0 \ = \ 1$}.
This results in a magnitude difference \mbox{$\Delta m \ = \ m - m_0$}, where $m_0$ is the 
integrated flux over the unspotted surface.
This procedure has been shown to yield rather accurate lightcurve fits \citep{Wolter2008}.

%\subsubsection{Line barycenters and light curves computed from DI reconstructions}
\subsubsection{Results}
\label{Sec:DIrvlc-res}

%We use the techniques described in Sect.~\ref{Sec:DIrvlc} to compute light curves and line barycenters from our Doppler images;
We present the lightcurves and line barycenters computed from our Doppler images in Fig.~\ref{Fig:DIRVLC}.
To assist discussion, we introduce phase intervals labeled 1 to 4.
Each interval represents a characteristic position of the dominant spot group on the visible stellar disk.
The appearance of the visible disk during these phase intervals is illustrated in Fig.~\ref{Fig:DI6439-phases}.

The RV curve of V889 Her, whose observation time interval encompasses that of our Doppler images,
is placed at the top of Fig.~\ref{Fig:DIRVLC}.
It was observed during a time span of roughly $500$~days, although most data points come from a limited
interval of about $150$~days (interval~(c) of Fig.~\ref{Fig:RVdata}).
It shows variations of the measured RV that appear largely stable over 
the selected time interval with a peak-to-peak amplitude of about $1 \, 000$~m/s.

All RV curves on the left side of Fig.~\ref{Fig:DIRVLC}  have a similar shape.
Their structure is similar to the one expected from an activity-induced Rossiter-McLaughlin effect 
(see Sects.~\ref{Sec:RVactivity} and~\ref{Sec:RVstable-regions}).
%This is even true for the barycenters derived from the lower quality Doppler images, e.g. line~6265~\AA \ or line~6462~\AA \ of nights~$5\, -\, 8$, which indicates a high robustness of our RV shift determination from line barycenters.

Interval~1, starting at about phase~0.6, marks the main increase in~$S$.
During this interval the spot region~B (containing the dominant feature) appears at the limb of the visible stellar disk.
This leads to a decreased absorption in the blue line wing, causing an apparent RV shift towards longer wavelengths.
This shift increases while the entire spotted region rotates onto the disk,
which is illustrated in panel~(1) of Fig.~\ref{Fig:DI6439-phases}, then decreases again and finally
drops to zero in interval~2, when the spot region~B is centered on the stellar disk (panel~(2)).

% Note that the line barycenters computed for the $6265$~\AA \ line are constantly shifted by about $-400$~m/s.
% This could be due to an unrecognized blend or a slightly erroneous determination of the line center prior
% to the DI. While this has no significant effect on the Doppler map, it shows up at the higher sensitivity
% of the line barycenter measurements.
% Actually, this could be used to determine a more precise line center. Here we only correct for this error by
% marking the average of all points of the $6265$~\AA \ by a horizontal dashed line. 

After interval~2, the spot region~B moves towards the other edge of the stellar disk
and influences the star's redshifted light.
The normalized absorption at higher wavelengths is reduced leading to
an apparent RV shift towards lower wavelengths. This shift reaches its maximum when region~B
is located at the `red' edge of the stellar disk.
When spot region~B finally moves off the visible disk, the line barycenter returns back to zero.

The previous discussion only covered the
influence of the large spot feature located at B, which dominates the shape of the RV curves.
However, other spotted regions influence the RV curve as well as can especially be seen in 
interval~4; the associated disk appearance is found in panel~(4) of Fig.~\ref{Fig:DI6439-phases}.
During this phase interval the dominant feature~B is on the backside of the star 
and does not influence our RV curve.
Here, spot group~D (cf. Fig.~\ref{Fig:DIimages}) dominates the behavior of the line barycenter as 
its influence is not masked by the much stronger contribution of spot group~B.
%Indeed, the largest change in shape of the RV~curves of Fig.~\ref{Fig:DIRVLC} between
%the \mbox{nights~$1\, -\, 4$} \mbox{and~$5\, -\, 8$} %Doppler images
%occurs in this interval~4: depending on the spectral line, a rising slope turns into a narrow plateau
%or a plateau changes into a shallow dip.

Summing up, the observed radial velocities (RV~curve) and the line barycenters ($S$~curve) 
show a far-reaching agreement in shape.
There is an apparent phase shift of about~0.1 by which the RV~curve lags behind the $S$~curves,
most pronounced in the position of the maximum. This apparent shift could be due to
small-scale variations of the spot pattern during the roughly \mbox{$100 \, -\, 200$~rotations}
covered by this interval of the RV~curve.
Alternatively, this shift could be caused by small or low-contrast spot groups not
reliably reconstructed in our Doppler maps.
Apart from this, the RV and $S$~curves even have comparable amplitudes.
%this agreement is poorer for the \mbox{$5\, -\, 8$~night} images 
%which yielded a less pronounced reconstruction of low-latitude features.

In addition to the line barycenters, the right column of Fig.~\ref{Fig:DIRVLC} contains the lightcurves 
computed from our Doppler maps.
The rotational evolution discussed above for the RV curves can be completely followed in the lightcurves;
zero passages of the former quite precisely coincide with extrema of the latter.
A more symmetric lightcurve is associated with a spot distribution in longitude that is also more symmetric.
Such a spot distribution, in turn, leads to a more symmetric RV~curve.
This can, for example, be seen when comparing the $6122$~\AA \ RV and lightcurves of nights~$1\, -\, 4$ and~$9\, -\, 12$, respectively. \\

% ------
% 
% $\longrightarrow$ Moved from section 5.3:
% 
% As discussed in Sect.~\ref{Sec:DIrvlc}, this curve agrees well with Doppler images obtained from data inside the same time interval.
% The modulation is due to an activity-induced Rossiter-McLaughlin effect caused by the stellar spot distribution and
% confirms that the global spot distribution of V889 Her was stable within at least one year.
% To our knowledge, this is the first time that activity-related RV measurements are successfully compared to reconstructed spot distributions,
% which shows that RV variations can be a powerful tool for stellar activity research as well.

%%%%%%%%%%%%%%%%%%%%%%%%%%%%%%%%%%%%%%%%%%%%%%%%%%%%%%%%%%%%%%
\section{Discussion}
\label{Sec:IntDis}

Lightcurves, Doppler images and RV measurements represent complementary data sets for the analysis of large-scale surface structures and their temporal evolution.
Signatures of activity can be compared and confirmed between them.
In accordance with previous authors, our observations show a significant fraction of V889 Her's surface covered with spots.
These are primarily concentrated in large regions near the poles and within two active regions at lower latitudes.

The analysis of the Hipparcos photometry (taken between $1990$ and $1993$) indicates that these active regions were confined to longitudes between 200\textdegree \ to 300\textdegree \ and 80\textdegree \ to 150\textdegree \ for more than two years.
Compared to our Doppler images and RV measurements, which show the spot distribution almost 10 years later, we find a qualitatively similar distribution shifted by at least several dozens of degrees in longitude.
The center of the dominant feature is located at about 20\textdegree \ longitude.
The second active region apparently had much less influence on the RV data; in the Doppler images, where only small signatures of other large spotted regions can be seen, it is located between 150\textdegree \ and 250\textdegree \ longitude.
From a global point of view, this is rather similar to the photometric results.

Concerning the inner structure and shape of active regions, which have a maximum diameter of about 30\textdegree \ to 60\textdegree, Doppler images yield information %that 2-spot models of light curves cannot contain.
our lightcurve models cannot contain.
The DI surface maps show asymmetrical and inhomogeneous active regions.
%The DI surface maps show asymmetrical, inhomogeneous spots with subdivisions inside of active regions and several individual spots outside of them.
%Quite naturally this is different compared to the 2-spot models where we initially assumed two circular, homogeneous `spots' because more detailed models are not constrained by the light curves.
These should not be confused with the active regions reconstructed in the lightcurve models where primarily information on longitudinal position, and no structure, is obtained.
Additionally the latter contain only information averaged over dozens of rotations.
This means that the size of an active region represents the temporally averaged area of the star with high activity; its `center' indicates the position with highest activity on average.
%The quality of our fit clearly indicates that the temporal changes are only minor compared to the depths of the minima and the errors of the data points (see Fig.~\ref{Fig:HippLightCurve} and Table~\ref{Tab:HippFitParErr}).
The quality of our fit clearly indicates that the temporal changes are only minor compared to the depths of the minima,
and we show that our models present statistically credible reconstructions of the photometry,
even though the poor phase sampling and the long time basis of the Hipparcos observations complicate the interpretation of this data set.
%Doppler images, reconstructed from data covering only a few rotations, can be seen as higher-resolution snapshots of what the 2-spot models support.

Previously, surface reconstructions of V889 Her had been derived by \citet{Strassmeier2003} and \citet{Marsden2006}.
Their main characteristics do not significantly differ from our Doppler images showing large polar spots down to almost 60\textdegree \ latitude and a few smaller spots at about 30\textdegree \ latitude.
The sizes are comparable as well and lie between about 20\textdegree \ to 30\textdegree ; the dominant feature of our reconstructions may even have a size of almost 60\textdegree \ in diameter (at least in longitudinal direction), although a possible multi-component structure cannot be ruled out.
A closer look even reveals some similarity to our lightcurve models. % 2-spot models.
The reconstructions of both authors show dominant surface features at about 300\textdegree \ longitude which roughly agrees with the location of one of our active regions.
Unfortunately this is very likely only a coincidence since
the long time distance between the different observations leads to phase errors of about 0.1 to 0.3 due to the rotation period's error (see Sect.~\ref{Sec:RVRP}).
This makes it hard to compare the spots' positions in different surface maps directly.
%% UW 2009feb - Streichen
%It is probably possible to identify similar distributions if relative positions are repeated in different surface maps, but from our point of view this is not reliably possible with the surface reconstructions of V889 Her published so far.

% Although \citet{Marsden2006} find differential rotation on this star, we do not have significant evidence either in favor of or against it in our data sets.
% The quality of our Doppler imaging data and reconstructions does not constrain differential rotation.
% The clear identification of features significantly shifted during the observed 9 rotations is hardly possible.
% %test do not lead to any statistical improvement of the reconstructed images.
% The interpretation of the lightcurves does not need the introduction of differential rotation, although it does not exclude it as well;
% statistically a rigid rotation model is sufficient.
% If, on the other hand, we assume the results of \citet{Marsden2006} also valid for out data set,
% %Using the results of \citeauthor{Marsden2006} in our interpretation would mean that the
% dominant spots are confined to latitudes of about $30$\textdegree.
%% UW 2009feb
While \citet{Marsden2006} observe differential rotation on this star, in our data sets 
we do not find any evidence in favor of or against it.
The quality of our Doppler imaging data and reconstructions does not allow to
significantly constrain differential rotation, mostly because of their lack of
well-defined and asymmetric high-latitude spots.
The interpretation of the lightcurves does not require the introduction of differential rotation, although it does not exclude it as well;
statistically a rigid rotation model is sufficient.
However, if strong differential rotation measured by \citet{Marsden2006} also 
applies to the time span of our RV and photometric data sets, this implies that the
dominant spots are confined to a rather narrow latitude band close to about $30$\textdegree \ latitude.

To compare the modulation of high-precision RV measurements to spot distributions, we present a method to determine RV shifts from DI line profile reconstructions.
In this process the shifted line center of an asymmetric profile is compared to the RV shift derived in standard RV measurement techniques, in our case from iodine cell spectra.
Our results show a striking agreement of this method %, especially for our most reliable Doppler images (night 1-4), 
with the conventional RV measurements, not only qualitatively but quantitatively as well.
This provides strong empirical evidence that both methods do not only produce comparable but equal results.
Their differences primarily derive from reconstruction errors of the DI line profiles.
An illustrative example of a complicated spot distribution causing the superposition of activity-induced Rossiter-McLaughlin effects is presented in this paper.
This is only possible due to the availability of Doppler images and contemporaneous high-precision RV measurements.

%%%%%%%%%%%%%%%%%%%%%%%%%%%%%%%%%%%%%%%%%%%%%%%%%%%%%%%%%
\section{Summary}
\label{Sec:Summary}

Using the example of the solar-type, fast rotating star V889~Her, we show that high-accuracy RV measurements can be used to study stellar activity, in particular large-scale spot distributions.

The modulation of V889~Her's RV curve with a period of
\mbox{$P \, = \, 1.3372 \, \pm \, 0.0004$~days} confirms \citeauthor{Strassmeier2003}'s rotation period of \mbox{$P \, = \, 1.3371 \, \pm \, 0.0002$~days}.
Our data are consistent with no differential rotation; if present, the spots must be confined to a limited range of latitudes.
%We cannot neglect or confirm differential rotation on the surface; if true, this rotation period belongs to dominant starspots confined to a limited latitude range.}
A large subinterval of the data yields a stable and characteristic RV curve due to an activity-induced Rossiter-McLaughlin effect.

This curve is compared to contemporaneous Doppler images.
RV shifts derived from different surface reconstructions match the RV observations well
confirming that the observed RV~modulation is caused by large-scale and long-term stable spotted regions.
This also shows that RV shifts can reliably be determined from line profiles reconstructed in the DI process.

Furthermore we confirm
the long-term stability of large-scale surface structures on V889~Her modeling the Hipparcos photometry.
%To this end we use a 2-spot modeling approach.
Pronounced lightcurve minima preserve their position almost unchanged throughout most of the Hipparcos observations.
They are associated with two large active regions located at approximately constant longitude but slowly 
changing size and latitude.

From our analysis we derive different time scales for structures of different sizes.
The evolution of `small scale' structures inside active regions is not resolved in our data, although the Doppler images indicate that spots of smaller sizes down to approximately 15\textdegree \ occur.
In some cases sizes and longitudes of active regions remain basically unchanged for up to 300~days.
The global configuration of a polar spot and two large, clearly separated spot groups
apparently existed for more than two years; these two spot groups even remained at largely constant longitudes.

Judging from the Doppler images of this paper and those of other authors (\citealt{Strassmeier2003}, \citealt{Marsden2006}), the polar spot is likely to be the most persistent surface feature with a lifetime of at least several years. 
This is also confirmed by the practically constant maximum brightness observed in the Hipparcos photometry.

%%%%%%%%%%%%%%%%%%%%%%%%%%%%%%%%%%%%%%%%%%%%%%%%%%%%%%%%%%%
\begin{acknowledgements}
We thank the NOT staff for their excellent support.
Many thanks to the services of the \textit{Vienna Atomic Line Database} (VALD).
This research has made use of the SIMBAD database, operated at CDS, Strasbourg, France; we appreciate their services very much.
K.H. and M.E. are members of the DFG Graduiertenkolleg 1351 \textit{Extrasolar Planets and their Host Stars}.
U.W. and S.C. acknowledge DLR support (50OR0105).
\end{acknowledgements}

%%%%%%%%%%%%%%%%%%%%%%%%%%%%%%%%%%%%%%%%%%%%%%%%%%%%%%%%%%%%%%

\bibliographystyle{aa}
\bibliography{referenz}

%%%%%%%%%%%%%%%%%%%%%%%%%%%%%%%%%%%%%%%%%%%%%%%%%%%%%%%%%%%%%%%
\appendix

% \section{Light curve model - parameter correlations}
% \label{App:LCerror}
% 
% The error determination for our light curve models is complex because the parameters may be correlated.
% This would mean that our results for the parameter uncertainties in Sect.~\ref{Sec:StableHIPP}, which are obtained with the method presented in Sect.~\ref{Sec:LCmethods}, are only lower limits. \\
% %
% Qualitative arguments for or against possible correlations are as follows.
% In our case the fit of both active regions is decoupled in good approximation since they are separated by typically 100\textdegree \ or more in longitude, which are well-determined by the light curve minima given sufficient depth and phase coverage.
% Correlations between spot longitudes and latitudes are unlikely because changes in one direction can hardly be compensated by changes in a `perpendicular' direction.
% There exists a correlation between latitude and size of a spot because broader light curves minima can be generated by high-latitude features and/or large spots as well.
% Higher positions lead to a decrease of the minimum's depth due to worse visibility, a large spot increases the minimum's depth and assists its broadening.
% The arguments of this section were confirmed by numerical simulations.
% Their presentation is beyond the scope of this paper, but detailed discussions on this topic and our 2-spot modeling approach are planned in forthcoming publications.

\section{Analytic proof of equation~\ref{Eq:bary}}
\label{App:Eq1}

In Sect.~\ref{Sec:DIrvlc} we introduce Eq.~\ref{Eq:bary} for the barycenter of a spectral line in analogy to
the barycenter determination of a system of masses
\mbox{$x_s \, = \, (\sum_i x_i \ m_i) / M$}, where \mbox{$M \, = \, \sum_i m_i$}.
We start with the continuous representation of Eq.~\ref{Eq:bary}, which reads
\begin{equation}
S \ = \ \frac{1}{W_{\mathrm{Eq}}} \int \limits_{-x_{\mathrm{max}}}^{+x_{\mathrm{max}}} x \ \Big[ 1-y(x-x_c) \Big] \ dx \ = \ x_c,
\label{Eq:AppBary}
\end{equation}
with $x$ denoting the radial velocity axis and $y(x)$ representing the normalized intensity ($0 \le y \le 1$).
$x_c$ is the center position of the line, which we show is equal to the barycenter $S$.
Note that the equivalent width, \mbox{$W_{\mathrm{Eq}} \, = \, \int (1-y(x_i)) \, d x$}, appears by full analogy with the total mass, $M$, here, and we presume it to be constant.
The line shape described by $y(x)$ can be arbitrary as long as the integral $\int (1-y(x_i)) \, d x$ exists;
however, the radial velocity axis must be chosen so that $\int x \, (1-y(x)) \, dx = 0$.

Now we aim at proving the correctness of the last equality in Eq.~\ref{Eq:AppBary}.
Note that with this definition \mbox{$(1-y(x))/W_{\mathrm{Eq}}$} becomes a distribution function; from the mathematical point of view the
integration boundaries may easily be extended to cover an arbitrarily large range, however, in the analysis of real data we
usually favor tight boundaries to avoid contamination. Therefore, we explicitly consider them here and, further, restrict ourselves
to small shifts, \mbox{$x_c \, \ll \, x_{\mathrm{max}}$}.

Recasting Eq.~\ref{Eq:AppBary} we obtain
\begin{align}
S  & \ = \ -\frac{1}{W_{\mathrm{Eq}}} \int \limits_{-x_{\mathrm{max}}}^{+x_{\mathrm{max}}} x \ y(x-x_c) \ dx \nonumber
\end{align}
with the substitution $a \, = \, x-x_c$
\begin{align}
S & \ = \ - \frac{1}{W_{\mathrm{Eq}}} \int \limits_{-x_{\mathrm{max}}-x_c}^{+x_{\mathrm{max}}-x_c} (a+x_c) \ y(a) \ da \notag \\
  & \ = \ - \frac{1}{W_{\mathrm{Eq}}} \int \limits_{-x_{\mathrm{max}}-x_c}^{+x_{\mathrm{max}}-x_c}
          \Big[ a \ y(a) \ + \ x_c \ y(a) \Big] \ da \ . \label{Eq:App2Int}
\end{align}
Above, $x_c$ represents an arbitrarily signed shift of the line's center, which we in the following define as positive without loss of generality.

We discuss the two terms of Eq.~\ref{Eq:App2Int} separately. The first can be treated as follows:
\begin{align*}
& - \frac{1}{W_{\mathrm{Eq}}} \int \limits_{-x_{\mathrm{max}}-x_c}^{+x_{\mathrm{max}}-x_c} a \ y(a) \ da \ = \\
& - \frac{1}{W_{\mathrm{Eq}}} \left(
\underbrace{\int \limits_{-x_{\mathrm{max}}-x_c}^{-x_{\mathrm{max}}+x_c} a \ y(a) \ da}_{\approx \ -2x_c x_{\mathrm{max}}}
+ \underbrace{\int \limits_{-x_{\mathrm{max}}+x_c}^{+x_{\mathrm{max}}-x_c} a \ y(a) \ da}_{= \ 0} \right) \ .
\end{align*}
In the approximation of the first integral we applied the relation \mbox{$y(a) \, \approx \, 1$}, which is valid within the integration interval ($x_{\mathrm{max}}$ must be chosen accordingly), and the second integral is zero by definition.

We resume with the second term of Eq.~\ref{Eq:App2Int}:
\begin{align*}
- \frac{1}{W_{\mathrm{Eq}}} \int \limits_{-x_{\mathrm{max}}-x_c}^{+x_{\mathrm{max}}-x_c} x_c \ y(a) \ da
& \ = \ - \frac{x_c}{W_{\mathrm{Eq}}} \left( 2 x_{\mathrm{max}} - W_{\mathrm{Eq}} \right) \\
& \ = \ - \frac{2 \, x_{\mathrm{max}} \, x_c}{W_{\mathrm{Eq}}} + x_c \ .
\end{align*}
For the last calculations we use
$$
W_{\mathrm{Eq}} \ \approx \int \limits_{-x_{\mathrm{max}}-x_c}^{+x_{\mathrm{max}}-x_c} \Big( 1 - y(a) \Big) \ da \ 
= \ 2 x_{\mathrm{max}} \, - \int \limits_{-x_{\mathrm{max}}-x_c}^{+x_{\mathrm{max}}-x_c} y(a) \ da \ .
$$

Inserting the result into Eq.~\ref{Eq:App2Int}, the equation
\begin{align}
S \ = \ \frac{2 \, x_c \, x_{\mathrm{max}}}{W_{\mathrm{Eq}}} - \frac{2 \, x_c \, x_{\mathrm{max}}}{W_{\mathrm{Eq}}} + x_c \ = \ x_c \ \
\mathrm{q.e.d.}
\end{align}
emerges, which is exactly the equality we strove to prove.

\section{Cross-correlation of Doppler images}
\label{App:DIcrosscor}

\begin{figure}[h]
  \centering
  \includegraphics[width=0.48\textwidth]{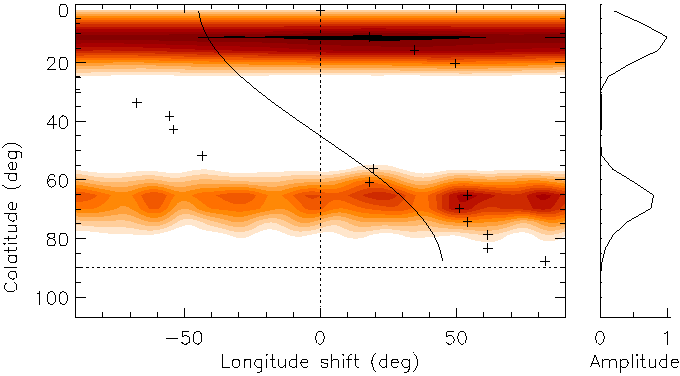} \\
  \includegraphics[width=0.48\textwidth]{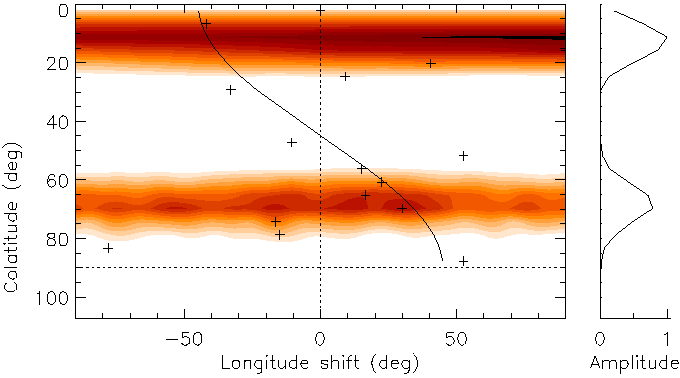}
  \caption{
  Cross-correlation maps comparing the Doppler images of Fig.~\ref{Fig:DIimages} (upper panel)
  and reconstructions using \citeauthor{Marsden2006}'s differential rotation law (lower panel).
  See text for detailed information.
  \label{Fig:DIcrosscor}}
\end{figure}

Figure~\ref{Fig:DIcrosscor} (upper panel) shows the cross-correlation map comparing the Doppler images of Fig.~\ref{Fig:DIimages},
i.e. the reconstructions for \mbox{nights~$1\,-\,4$} and \mbox{$9\,-\,12$} using line~$6122$ and rigid rotation.
For each co-latitude the map shows the cross-correlation, normalized to the overall maximum,
obtained when shifting the spots of map \mbox{$9\,-\,12$} by the given longitude angle.
Darker orange shades indicate larger correlation values.
Crosses mark the correlation maximum for each co-latitude, their values are shown in the right-hand graph,
also illustrating the assignment of orange hues.
For comparison, the smooth, sine-like curve renders the surface shear resulting from \citeauthor{Marsden2006}'s differential rotation law.

Figure~\ref{Fig:DIcrosscor} (lower panel) shows the same for Doppler images reconstructed adopting
\citeauthor{Marsden2006}'s differential rotation law.
The correlation maps do not support or disprove any specific rotation law.

\begin{figure*}
  \centering
  \rotatebox{90}{\hspace*{0.15\textheight} Line 6439~\AA, Nights~$1\, -\, 4$}
%   \rotatebox{90}{
     \includegraphics[height=0.46\textheight]{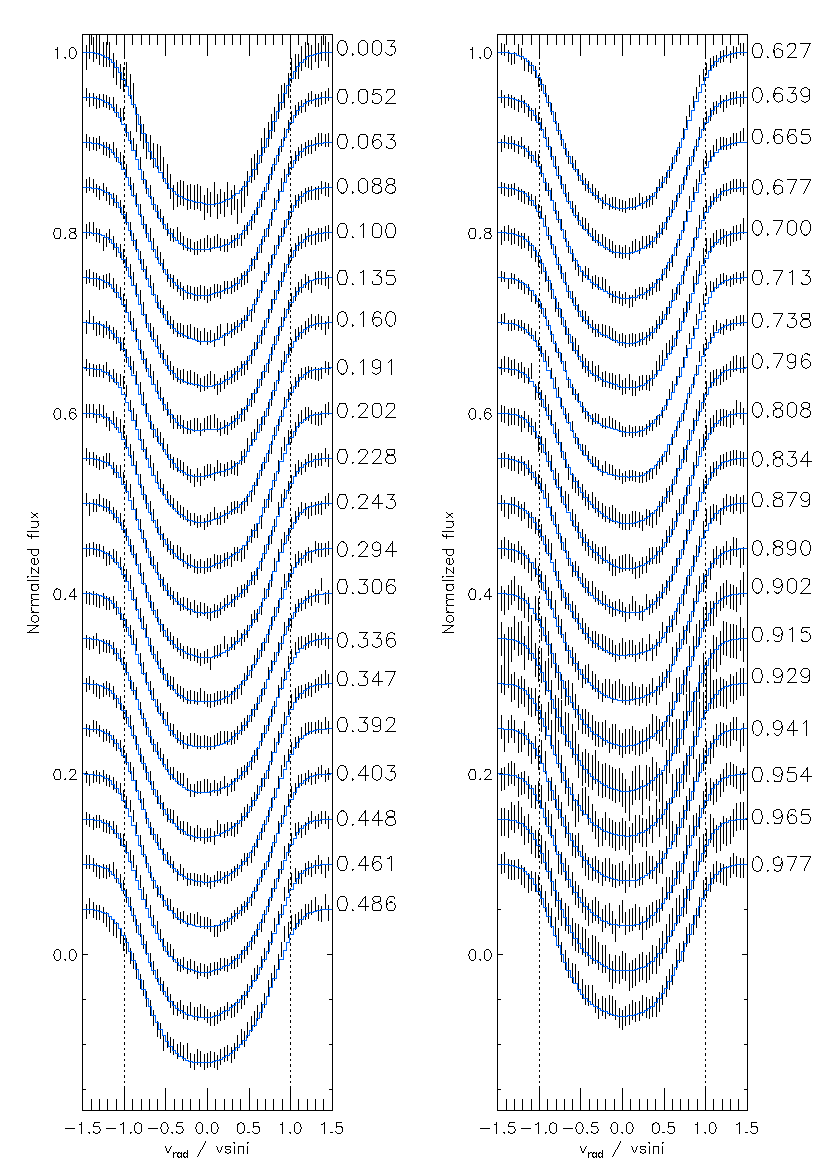} \vrule \vspace*{1pt} \vrule \vspace*{4pt}
  \rotatebox{90}{\hspace*{0.15\textheight} Line 6439~\AA, Nights~$9\, -\, 12$}
     \includegraphics[height=0.46\textheight]{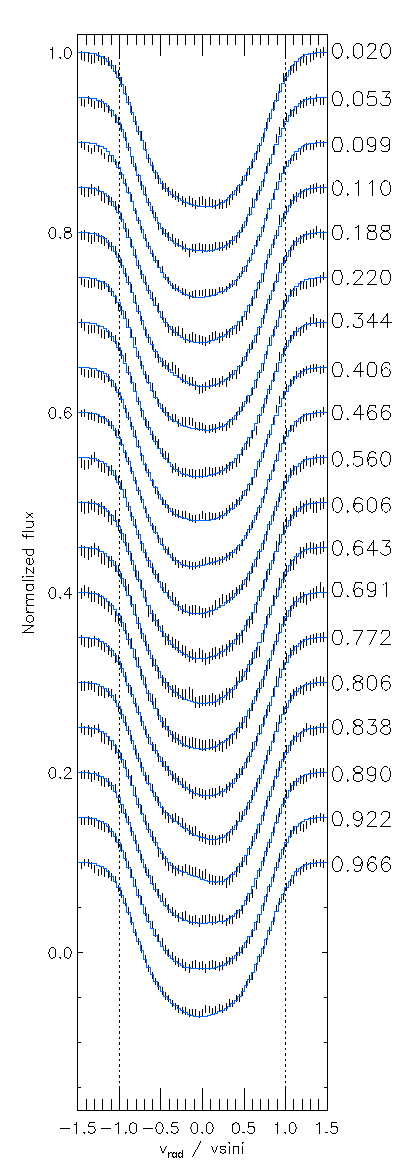} \\
%   }
% \caption{
% Stack plot of all DI line profile reconstructions of line 6439~\AA \ from \mbox{nights~$1\, -\, 4$} (two lower panel, 39 pases)
% and \mbox{nights~$5\, -\, 8$} (upper panel, 19 phases).
% See App.~\ref{App:DIlines} for details.
% \label{Fig:DI-lineprofiles6439}}
%\end{figure*}
%\hrule
%\begin{figure*}
  \centering
  \rotatebox{90}{\hspace*{0.15\textheight} Line 6122~\AA, Nights~$1\, -\, 4$}
%   \rotatebox{90}{
     \includegraphics[height=0.46\textheight]{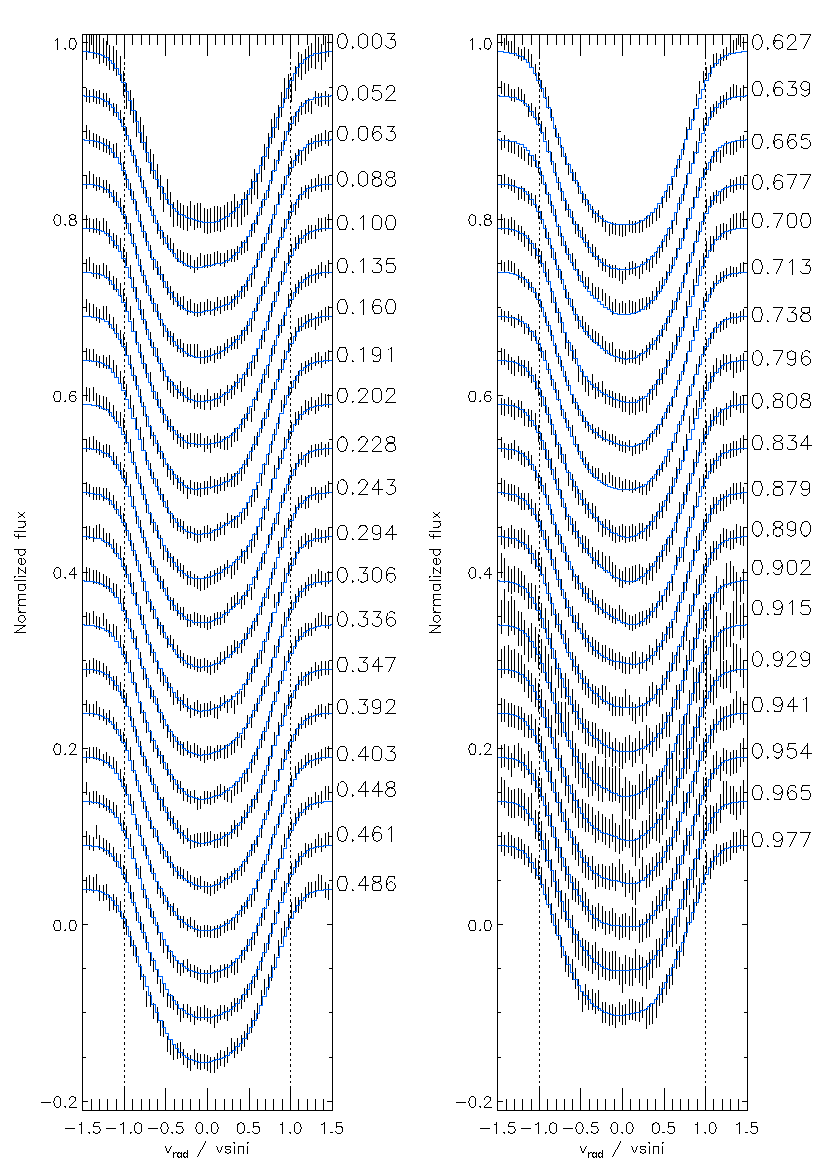} \vrule \vspace*{1pt} \vrule \vspace*{4pt}
  \rotatebox{90}{\hspace*{0.15\textheight} Line 6122~\AA, Nights~$9\, -\, 12$}
     \includegraphics[height=0.46\textheight]{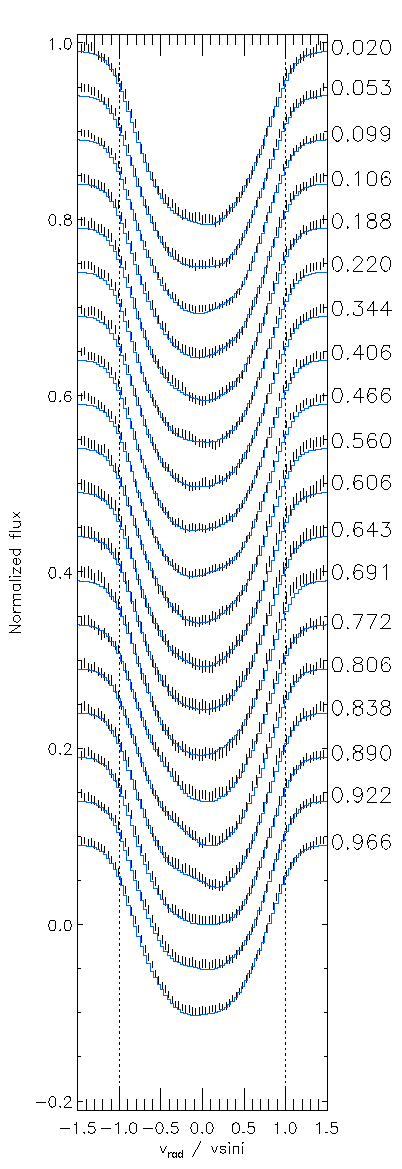}
%   }
\caption{
Stack plot of all DI line profile reconstructions.
The left column shows the $39$~phases of \mbox{nights~$1\, -\, 4$}, the right column the $19$~phases of the
\mbox{nights~$9\, -\, 12$} observations.
The two upper panels show line 6439~\AA, the two lower panels show 6122~\AA.
%of line 6122~\AA \ from \mbox{nights~$1\, -\, 4$} (two lower panel, 39 pases)
%and \mbox{nights~$5\, -\, 8$} (upper panel, 19 phases).
See App.~\ref{App:DIlines} for more details.
\label{Fig:DI-lineprofiles}}
\end{figure*}

\section{DI line profile reconstructions}
\label{App:DIlines}

We give the DI line profile reconstructions of all Doppler images presented in Sect.~\ref{Sec:DIimages}.
Figure~\ref{Fig:DI-lineprofiles} contains the profiles of the 6439~\AA \ (top panels)
and the 6122~\AA  (bottom panels) Ca~I lines, 39 phases for \mbox{nights~$1\, -\, 4$} (left column)
and 19 phases for \mbox{nights~$9\, -\, 12$} (right column).
%Figure~\ref{Fig:DI-lineprofiles6122} gives the same for spectral line 6122~\AA.
All lines are continuum normalized but shifted in flux by $0.05$ (stacked plot).
Rotation phases increase from top to bottom; they are given at the right border.
The observed line profiles are shown as vertical lines indicating their observation errors,
the DI reconstructions (blue) are overplotted with a straight line.
Both are plotted over rotation velocity $v_{\mathrm{rad}}$ in units of $v \sin(i)$.

% \newpage
% 
% \begin{figure*}
%  \includegraphics[clip=,bb=0 300 800 700, scale=.6]{print.pdf}
%  \includegraphics[clip=,bb=0 300 800 700, scale=.6]{print2.pdf}
%  \caption{
% The plots illustrate the results of our cross-correlation analysis, comparing Doppler images reconstructed from the line profiles of nights 1-4 and 5-8, respectively. For the compared DIs we used the 6122\AA line profiles and Marsden et al.'s differential rotation law (Peq = 1.313 days, alpha = delta Omega / Omega\_equator = 0.084). \\
% Second plot: \\
% It shows the colour coded cross-correlation amplitude for each colatitude as a function of shear angle. The crosses indicate the maximum for each individual colatitude. The sine-like graph shows the expected shear after 4 nights (3 stellar roations), assuming Marsden et al.'s rotation parameters and a $\sin^2$ law. \\
% It is evident, that the correlation maxima of our DI do not follow the Marsden et al. shear law. \\
% First plot: \\
% It shows 1d-slices through the second plot and illustrates that these maxima are largely not significant. Above 25 deg colatitude, the correlation function is dominated by a broad maximum (or just a slope) around some poorly defined positive shear angle (red curve). At intermediate colatitudes there are practically no features in our DIs leading to a largely zero correlation function (green graph). \\
% Finally, closer to the equator,  the correlation function exhibits several local maxima of comparable amplitude (blue graph).
% }
% \end{figure*}
% 
% 

\end{document}